\documentclass[a4paper,11pt]{JHEP}
\pdfoutput=1
\usepackage{graphicx}
\usepackage{amsmath}
\usepackage{amssymb}
\usepackage[T1]{fontenc}
\usepackage{multirow}

\newcommand{\mnras}{Mon. Not. R. Astron. Soc.}
\newcommand{\aap}{Astron.Astrophys.}

\preprint{CERN-PH-TH/2012-216, LAPTH-038/12}

\title{CMB photons shedding light on dark matter}

\author{
Ga\"elle Giesen$^a$,
Julien Lesgourgues$^{a,b,c}$,
Benjamin Audren$^a$, 
Yacine Ali-Ha\"{\i}moud$^{d}$\vspace{.2cm}\\ 
{$^a$}Institut de Th\'eorie des Ph\'enom\`enes Physiques,\\ \'Ecole Polytechnique
F\'ed\'erale de Lausanne,\\ CH-1015, Lausanne,
Switzerland.\vspace{.2cm}\\ 
{$^b$} CERN, Theory Division,\\ CH-1211
Geneva 23, Switzerland.\vspace{.2cm}\\ {$^c$} LAPTh (CNRS -
Universit\'e de Savoie), BP 110,\\ F-74941 Annecy-le-Vieux Cedex,
France.\vspace{.2cm}\\ {$^d$} Institute for Advanced Study,\\
Einstein Drive, Princeton, New Jersey 08540, USA}

\abstract{
The annihilation or decay of Dark Matter (DM) particles could affect the
thermal history of the universe and leave an observable signature in Cosmic
Microwave Background (CMB) anisotropies. We update constraints on the
annihilation rate of DM particles in the smooth cosmological background, using WMAP7
and recent small-scale CMB data. With a systematic analysis based on the Press-Schechter formalism, we also show that DM
annihilation in halos at small redshift may explain entirely 
the reionization patterns observed in the CMB, 
under reasonable assumptions concerning the concentration and
formation redshift of halos. We find that a mixed reionization model based on
DM annihilation in halos as well as star formation at a redshift $z\simeq 6.5$
could simultaneously account for CMB observations and satisfy constraints inferred
from the Gunn-Peterson effect. However, these models tend to reheat the inter-galactic medium (IGM) well above observational bounds: by including a realistic  prior on the IGM temperature at low redshift
and allowing most of the reionization to be due to star formation, we find stronger cosmological bounds on the annihilation cross-section than with the CMB alone.
}


\begin{document}

\section{Introduction}

In the minimal $\Lambda$CDM model, the CMB has very little to say about Dark
Matter (DM), apart from a measurement of the relic abundance parameter
$\Omega_{\rm DM}h^2$. However, there is a chance that DM could leave another signature in the CMB. In the case of annihilating DM, if the ratio of the annihilation cross section over the mass is not too
small, annihilation products could contribute to the ionization of the thermal bath,
and affect the history of recombination and reionization. This has
already been discussed in detail in several references including \cite{Padmanabhan:2005es,Zhang:2006fr,Natarajan:2008pk,Belikov2009,Cirelli2009,Galli2009,Slatyer2009,Natarajan2010,Galli2011,Finkbeiner2011,Hutsi:2011vx,Natarajan2012}. In the case of
decaying dark matter, similar effects could take place if the particle lifetime
is not too large \cite{Chen:2003gz,Ichiki:2004vi,Zhang:2007zzh,Kasuya:2006fq,Yeung:2012ya}. 

In a detailed study of these mechanisms, assumptions concerning the nature of dark matter are of course crucial. Different DM particles may annihilate or decay in different channels (into hadrons, leptons, gauge bosons, etc.). The produced particles can themselves decay in several steps, until only stable particles like electrons, photons and neutrinos remain. While the energy contained in neutrinos is lost for the cosmic plasma, other decay products can contribute to the ionization and heating of this plasma. The authors of \cite{Slatyer2009} computed $f(z)$, the redshift-dependent fraction of the energy produced by annihilations that contributes to the ionization and heating of the plasma, for several WIMP models.  Similar calculations could be carried on any type of annihilating or decaying DM particles. In a cosmological analysis like that presented in this paper, we don't need to refer specifically to a given DM particle, and we may incorporate all model-dependent particle physics assumptions in the unknown fraction $f(z)$.

In this work, we revisit the impact on the CMB of annihilating DM.  
We will not introduce any new physical ingredients, neither from the point of view of particle physics, nor from that of structure formation models. However, we will present some new generic parametrizations of the relevant phenomena, in order to perform a systematic comparison of various models to recent CMB data sets. This will allow us to tighten some of the bounds presented previously in the literature.

In section 2, we summarize the impact of DM annihilation on recombination, and
explain how we took it into account by modifying the public Boltzmann code {\sc
class}. This code can simulate recombination with either of the two public
codes {\sc recfast} and {\sc hyrec}. We double-checked our results by modifying
the two algorithms. We show that they give the same results, but for
non-trivial models like those assumed in section 4, only {\sc hyrec} remains
numerically stable.

In section 3, we focus on the effects of annihilation in the smooth DM background
distribution. We find similar but slightly stronger bounds than in recent
studies, thanks to our updated CMB data set. With a generic parametrization of
the redshift-dependence of the function accounting for the fraction of energy
released to the gas, we confirm that current CMB data is not sensitive to this
dependence.

In section 4, we consider the additional effect of enhanced DM annihilation in
halos at small redshifts. This effect has been previously discussed in several
references including~\cite{Natarajan:2008pk,Belikov2009,Cirelli2009,Natarajan2010,Hutsi:2011vx,Natarajan2012}. Some of these works suggest
that it could account for a significant fraction (if not the totality) of the reionization of the universe at low redshift. We derive an
approximate but rather generic parametrization of this effect, and carry on the
first systematic parameter inference using current CMB data in a model with
reionization from annihilation. We also confront to the data a mixed model,
with reionization explained both by DM annihilation and star formation, and
discuss the relevance of this model for explaining simultaneously CMB data and
Gunn-Peterson bounds. Finally, we show that these models tend to reheat the inter-galactic medium (IGM) well above observational bounds; by including a realistic  prior on the IGM temperature at low redshift
and allowing most of the reionization to be due to star formation, we find stronger cosmological bounds on the annihilation cross-section than with the CMB alone.

Our general conclusions and future directions of research are outlined in section 5.

%%%%%%%%%%%%%%%%%%%%%%%%%%%%%%%%%%%%%%%%%%%%%
\section{Equations of recombination with Dark Matter annihilation}

\subsection{Standard recombination}

Before discussing the impact of DM annihilation (or alternatively dark matter
decay) on the recombination history, we first briefly review the
standard recombination model. We start by describing the simple three-level atom model of Peebles
\cite{Peebles1968, Zeldovich:1969} and then discuss the subsequent improvements
of this model.

In what follows we shall assume that helium has entirely recombined (which
is indeed the case for all redshifts of interest) and only deal with
hydrogen recombination. We denote $n_e$ the number density of free electrons, $n_H$
the total number density of hydrogen (in ionized and atomic forms),
$x_e=\frac{n_e}{n_H}$ the free electron fraction and $T_{\rm M}$
(resp. $T_r$) the matter (resp. photon) temperature.

\subsubsection{The effective three-level atom (TLA) model}

It is well known
that direct recombinations to the ground state are highly inefficient:
if a hydrogen atom forms directly in its ground state, it emits a
photon which is going to immediately ionize an
other atom, leaving the overall free electron fraction unchanged. The basic
idea of Peebles' ``case B recombination'' is that efficient
recombination only takes place when the electron gets first captured
into an excited state $n \geq 2$, from which it cascades down to $n =
2$. The newly formed atom may then eventually reach the ground state, either by emitting a Lyman-$\alpha$
photon from $2p$, or by the $2s\rightarrow 1s$ two-photon process. The Lyman-$\alpha$ line being very optically thick, the net rate
of $2p\rightarrow 1s$ transitions is, to a first approximation, the rate at which Lyman-$\alpha$
photons redshift across the resonance. At early times ($ z\gtrsim
900$), the net rate of transitions from the $n=2$ state to the ground
state is much smaller than the rate at which excited atoms are
photoionized by CMB photons, and the slow $2 \rightarrow 1$ transitions
constitute the bottleneck of the recombination process. At late times
($z \lesssim 900$), the intensity of the radiation field drops, and
atoms that do recombine to an excited state almost certainly reach
the ground state; during this period the new bottleneck is the rate at
which free electrons and protons can encounter each other and recombine.

To correctly describe recombination, accounting for the effects
mentioned above, Peebles introduced the pre-factor $C$ defined as
\begin{equation}
C=\frac{1+K_H\Lambda_H n_H(1-x_e)}{1+K_H(\Lambda_H+\beta_H)n_H(1-x_e)}, 
\label{Cfactor}
\end{equation}
where $\Lambda_H=8.22458$ s$^{-1}$ is the decay rate of the $2s$
level, $K_H=\frac{\lambda_{\rm Ly \alpha}^3}{8 \pi H(z)}$ accounts for the
cosmological redshifting of Lyman-$\alpha$ photons\footnote{$[K_H
n_H (1-x_e)]^{-1}$ is the rate of escape of Lyman-$\alpha$ per atom in
the $2s$ state}, and $\beta_H$ is
the effective photoionization rate from $n=2$ (per atom in the $2s$
state). $C$ represents the probability for an electron in the $n=2$
state to get to the ground state before being ionized. The evolution equation for the free-electron fraction is then given by
\begin{equation}
\frac{d x_e}{dz}=\frac{1}{(1+z)H}C \left[ \alpha_H x_e^2 n_H - \beta_H
  (1- x_e) e^{-\frac{ h \nu_{\alpha}}{k_b T_{r}}}\right], \label{xe}
\end{equation}
where $\alpha_H$ is the case-B recombination coefficient and
$\nu_{\alpha}$ is the Lyman-$\alpha$ frequency. Because the effective
recombination rate per free electron $C \alpha_H x_e n_H$ is always
much smaller than the Hubble rate (due to the two bottlenecks
mentioned above), primordial recombination proceeds much slower than in Saha equilibrium.

In addition, the matter temperature is determined from the
Compton evolution equation:
\begin{align}
\frac{dT_{\rm M}}{dz}&=\frac{1}{(1+z)H} \frac{8\sigma_T a_r T_r^4}{3m_e c}\frac{x_e}{1+f_{He}+x_e}(T_{\rm M}-T_r)+\frac{2T_{\rm M}}{1+z}.
\label{T_M}\\
&= \frac{1}{(1+z)}\left[2 T_{\rm M} +\gamma(T_{\rm M} -T_r) \right],
\end{align}
where we have defined the dimensionless parameter 
\begin{equation*}
\gamma\equiv\frac{8 \sigma_T a_r T_r^4}{3 H m_e c}\frac{x_e}{1+f_{He}+x_e},
\end{equation*}
where $\sigma_T$ is the Thomson cross-section, $a_r$ the
radiation constant, $m_e$ the electron mass, $c$ the speed of
light and $f_{He}$ the fraction of helium by number of nuclei. When $\gamma \gg 1$, the matter temperature is locked to the radiation
temperature by Compton
heating, $T_{\rm M} \approx T_r \propto (1+z)$; it decays adiabatically as $T_{\rm M} \propto (1+z)^2$ when $\gamma \ll 1$.

\subsubsection{Beyond the TLA model}

With the prospect of upcoming high-precision data from the Planck
satellite, several groups have revisited the simple TLA model
presented above and introduced important corrections. Here we use the codes {\sc {\sc recfast}} \cite{Seager1999} and
{\sc hyrec} \cite{AliHaimoud2010} which implement these corrections, approximately
for the former and exactly for the latter. The corrections are of two
types:

$\bullet$ Highly excited states of hydrogen are not in Boltzmann
equilibrium with each other, and one must account for all bound-bound
and bound-free transitions involving them, including stimulated transitions. The original  {\sc {\sc recfast}} code accounted for these transitions approximately by
multiplying the case-B recombination coefficient (and effective
photoionization rate) by a ``fudge factor'' $F = 1.14$, fitted
to reproduce multilevel computations \cite{Seager1999}. The original computation of
Seager et al assumed that angular momentum substates were in
statistical equilibrium. This approximation, however, was shown not to be
accurate enough \cite{Grin2009, Chluba2010}. The latest
version of  {\sc {\sc recfast}}\footnote{This work was completed
  using {\sc recfast} v1.5.1. The next version 1.5.2, including new
  fudge factors leading to very good agreement with HyRec, was
  released after the submission of this paper.} attempts to account for these more detailed high-$n$
computations approximately by using a new fudge factor $F = 1.125$. 

It turns out that the effect of highly-excited states can be
\emph{exactly} and efficiently accounted for by generalizing the
case-B coefficient to a non-zero CMB temperature
\cite{AliHaimoud2010a, Burgin2010}. The code {\sc {\sc hyrec}} is
using precomputed effective recombination coefficients in an effective
few-level atom model, at virtually no speed cost compared to the TLA
model, and with the advantage of being exact.

$\bullet$ Being the recombination bottleneck at early times, the
Lyman-$\alpha$ escape and two-photon $2s\rightarrow 1s$ decays need to
be modeled very precisely. Several radiative transfer effects were
shown to be important for high-accuracy predictions of CMB
anisotropies (see for example Refs.~\cite{Kholupenko2006, Chluba2008,
  Hirata2008, Hirata2009, Chluba2010} and references
therein). Detailed codes such as {\sc hyrec} and {\sc cosmorec}
\cite{Chluba2010} account for all important radiative transfer effects
exactly, by evolving the radiation field numerically. This part of the
calculation is heavier computationally, but efficient implementations
render the runtime for the recombination calculation comparable
with the runtime of the Boltzmann code itself. {\sc recfast} accounts
for radiative transfer effects by adding a correction function to the
recombination rate $\dot{x}_e|_{\rm corr}$, fitted to reproduce the
detailed codes for cosmologies close to the current best-fit
value. 

\subsection{Parametrization of Dark Matter annihilation}

We wish to express the rate at which the energy released by DM annihilations is
injected in the thermal bath. In the next subsection, we will summarize how
this energy is used and in which proportions. 

We can write the energy injected into the plasma per unit of volume and time as the product of the number of DM particle pairs
$n_{pairs}$, the annihilation probability per unit of time $P_{\rm ann}$, the
released energy per annihilation $E_{\rm ann}$, and the redshift-dependent
fraction of released energy $f(z)$ absorbed by the gas
\begin{align}
\left.\frac{dE}{dV dt}\right|_{\rm DM} \!\! \!\!(z)&= n_{pairs}\cdotp P_{\rm ann} \cdotp E_{\rm ann}\cdotp f(z) =\frac{n_{\rm DM}}{2} \cdotp \langle\sigma v\rangle \cdotp n_{\rm DM}\cdotp 2 m_{\rm DM} c^2 \cdotp f(z) \nonumber
\\ &=\rho_{\rm c}^2 c^2 \Omega_{\rm DM}^{2}(1+z)^6 f(z)\frac{\langle \sigma v \rangle}{m_{\rm DM}}.
\label{DM_energy_rate}
\end{align}
In Eq.~(\ref{DM_energy_rate}), $\sigma$ is the annihilation cross-section, $v$ is the relative
velocity of DM particles, $\langle \sigma v \rangle$ is the average of
$\sigma \times v$ over the velocity distribution, $m_{\rm DM}$ the
mass of the DM particle, $\rho_{\rm c}=\frac{3 H_0^2}{8\pi G}$ the critical
density of the universe today (with $H_0$ the Hubble constant today),
and $\Omega_{\rm DM}$ the Dark Matter abundance today relative to the
critical density. 
In the case where DM consists of Dirac Fermions, there should be an extra factor 1/2 in the last two equalities (since only half of the pairs are made of one particle and one anti-particle). If this is the case, we can decide to absorb this factor in a redefinition of $f(z)$. Then, for a given cosmological evolution, all the model-dependent part of the energy injection rate can be parametrized by the
following function of redshift,
\begin{equation}
p_{\rm ann}(z)=f(z)\frac{\langle \sigma v\rangle}{m_{\rm DM}}.
\end{equation}
The authors of \cite{Slatyer2009} computed $f(z)$ for several WIMP models. They found that $f(z)$ is a smoothly decreasing  function, lying in the range from 0.2 to 0.9 at redshift 2500 (depending on the WIMP mass and dominant annihilation channel), and decreasing by a factor 2 to 5 at small redshift. Similar calculations could be carried for any type of annihilating or decaying DM particles.

\subsection{Effects of Dark Matter annihilation on the thermal history of the universe\label{DM_effects}}

The energy injected by DM annihilation has three effects: ionizing the
plasma, exciting hydrogen atoms 
, and heating the plasma \cite{Peebles2000,Galli2009}. A fraction of the atoms excited by the second mechanism will be subsequently ionized by CMB photons. Hence, the first two effects (illustrated in Figure~\ref{schema}) have a direct impact on the free electron fraction, and the last one on the matter temperature.
\FIGURE{
\includegraphics[scale=0.3]{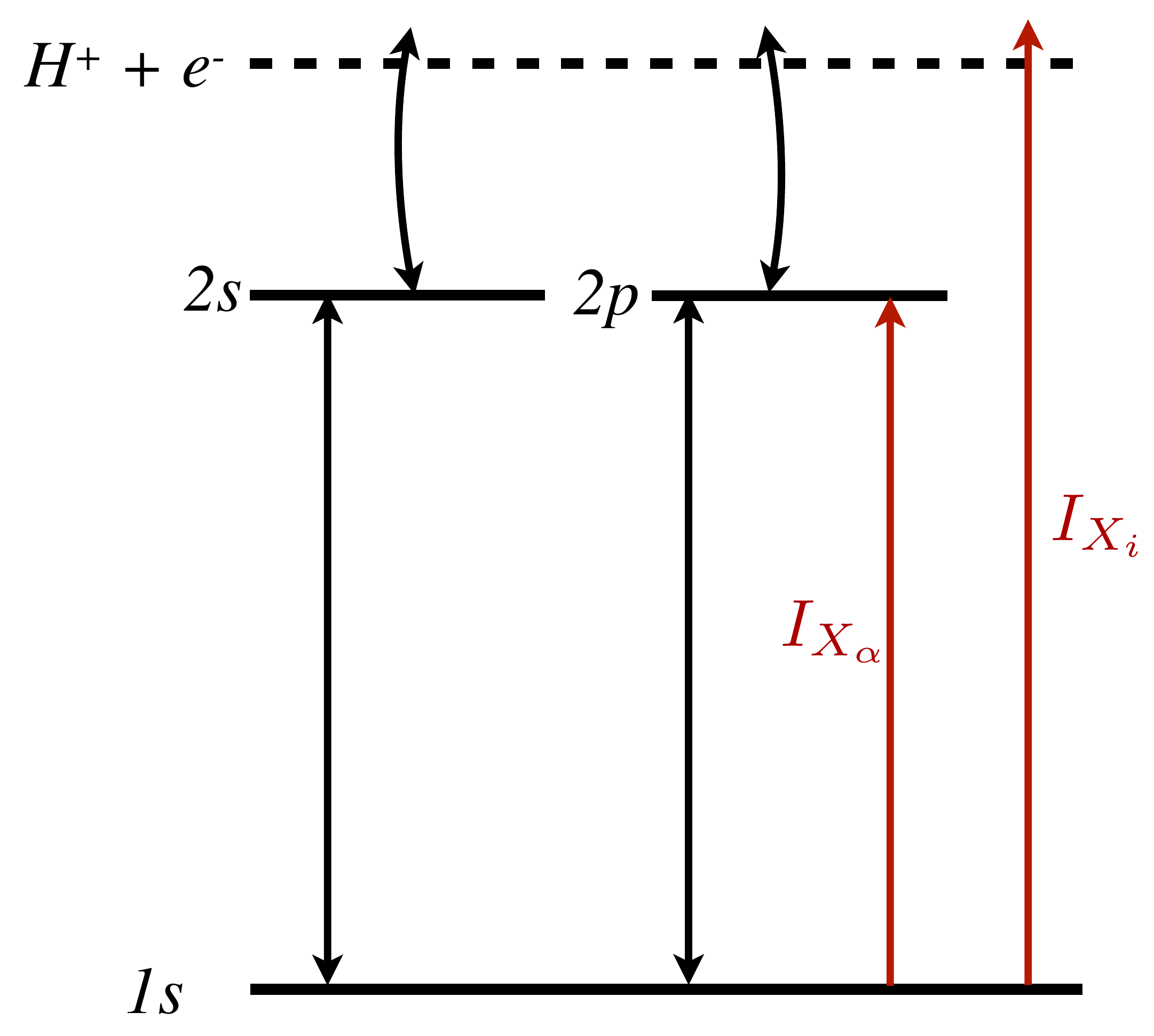}
\caption{Illustration of the impact of DM annihilation on Peeble's ``case B recombination'' model.\label{schema}}}

For simplicity, let us introduce the basic equations describing the
three effects of DM annihilation using the on-the-spot approximation,
which supposes that all interactions between the decay products of DM
annihilation and the plasma take place locally, on a time scale
negligible with respect to the expansion time scale.

\noindent \textbf{Ionization of the plasma.} 
In the on-the-spot approximation, if $\chi_i(z)$ denotes the fraction of the injected energy going into ionization, and $E_i$ the average ionization energy per baryon,
the number of direct ionizations per interval of redshift $dz$ reads
\begin{align}
I_{X_i}(z)=&-\frac{\chi_i(z)}{(1+z) \, H(z) \, n_H(z) \, E_i} \left.\frac{dE}{dV dt}\right|_{\rm DM},
\label{ionization}
\end{align}
where we used $dz/dt=-(1+z)H$.
Shull and Van Steenberg found that for a neutral gas, approximately 1/3 of the energy goes
into ionization \cite{Shull1982}. For an ionized gas, none of the energy can be
used for ionization. Thus, for a partially ionized gas, Chen and Kamionkowski
proposed to approximate $\chi_i$ by $(1-x_e)/{3}$ \cite{Chen:2003gz}.  
The fact that $\chi_i \propto (1- x_e)$ makes physical sense,
since the ionization rate must be proportional to the abundance of
neutral hydrogen.
\\

\noindent \textbf{Excitation of hydrogen.} 
The rate of collisional excitation of hydrogen (1s$\rightarrow$2p and
1s$\rightarrow$2s transitions, etc.) due to DM annihilation is similar
to that of direct ionization, with $E_i$ replaced by the
Lyman-$\alpha$ energy  $E_\alpha$ and $\chi_i(z)$ by the fraction
$\chi_\alpha(z)$ of the injected energy going into
excitations. 
Once a given atom is in the $n=2$ state, it has a probability $(1-C(z))$ to be ionized by CMB photons. Thus, the net ionization
rate per redshift interval $dz$ due to collisional excitations  
by DM annihilation products reads
\begin{align}
I_{X_\alpha}(z)=&-\frac{(1-C(z)) \, \chi_\alpha(z)}{(1+z) \, H(z) \, n_H(z) \, E_\alpha} \left.\frac{dE}{dV dt}\right|_{\rm DM} .
\label{excitation}
\end{align}
Chen and Kamionkowski showed that in first approximation one may assume $\chi_i = \chi_\alpha = (1-x_e)/{3}$. 
Note that this process is subdominant with respect to the direct ionization of the plasma. \\

\noindent \textbf{Heating of the plasma.} 
Finally, DM matter annihilation heats the plasma at a rate (per unit
of time) 
\begin{align}
\frac{d T_{\rm M}}{d t}\Big{|}_{\rm DM}=\frac{2 \chi_{\rm h}}{3 k_b n_H (1+f_{He}+x_e)}\left.\frac{dE}{dV dt}\right|_{\rm DM},
\label{heating}
\end{align}
with $\chi_{\rm h}=1-\chi_i-\chi_\alpha=(1+2x_e)/{3}$ the remaining fraction of the total injected energy. 

\vspace{0.3cm} 

The range of validity of these equations extends beyond the on-the-spot
approximation, provided that the ratio $\left. \frac{dE}{dV dt}
\right|_{\rm DM}$ stands for the effective injection rate at redshift
$z$ coming from DM annihilation at all redhsifts $z'\geq z$, taking into account
energy transfer and absorption processes between $z'$ and $z$. The function $f(z)$ was actually computed by \cite{Slatyer2009,Huetsi:2009ex,Hutsi:2011vx} beyond the on-the-spot approximation.

\subsection{Recombination equations with DM annihilation}

We can now write the modifications needed for each of the two recombination
codes {\sc {\sc recfast}} and {\sc hyrec}, both implemented in the Boltzmann
code {\sc class}\footnote{\tt http://class-code.net}
\cite{Lesgourgues:2011re,Blas:2011rf} used throughout this work. The point of
using two different codes is to compare the results and check that our approach
for including annihilation effects is robust and consistent. In addition, we
will see that in some of the cases discussed below, the second code is more
stable numerically and allows to explore more general models.

Given
equations~(\ref{DM_energy_rate} -- \ref{heating}), implementing DM annihilation
in the two codes only requires to add two new terms proportional to $p_{\rm ann}(z)$
in the basic equations for hydrogen recombination and matter temperature:
\begin{align}
\frac{d x_e}{dz}=&\left.\frac{d x_e}{dz}\right|_{\rm st} -
\frac{\rho_{\rm c}^2 c^2 \Omega^2_{\rm DM} (1+z)^5}{H(z)}  \left[ \frac{1-x_e(z)}{3 \,n_H(z)} p_{\rm ann}(z) \left(\frac{1}{E_i}+\frac{1-C(z)}{E_\alpha}\right)\right], \label{eq_recfast_1}\\
\frac{dT_{\rm M}}{dz}=&\left. \frac{dT_{\rm M}}{dz} \right|_{\rm st} - 
\frac{\rho_{\rm c}^2 c^2 \Omega^2_{\rm DM} (1+z)^5}{H(z)}
\left[\frac{2}{3 k_b}\frac{1+2x_e(z)}{3 \, n_H(z)}\frac{1}{1+f_{He}+x_e(z)}  p_{\rm ann}(z)\right],\label{eq_recfast_2}
\end{align}
where the subscript ``st'' stands for the standard rates, given by eqs~(\ref{xe},\ref{T_M}) for case B recombination. These equations neglect the
possibility that  a fraction of the energy released by DM annihilation would
serve for helium ionization. As in ref.~\cite{Galli2011}, we checked that such
a refinement would have a negligible impact on the CMB spectra. 

In {\sc hyrec}, it is also
necessary to write separately the impact of DM annihilation in the equations
accounting for approximation schemes, like the steady-state
approximation for the matter temperature at early times. In
appendix A, we write explicitly our modified {\sc hyrec} equations.
\\

In section~\ref{sec:halos}, we will introduce extra modifications allowing to account for DM annihilation at small redshift beyond the on-the-spot approximation.
Our modification to {\sc class}, including those in {\sc recfast} and {\sc
hyrec}, will be part of the next public distribution 1.5 of the code.

%%%%%%%%%%%%%%%%%%%%%%%%%%%%%%%%%%%%%%%%%%%%%%%%
\section{Dark Matter annihilation before structure formation and reionization}
\label{sec_const}

In this section, we wish to better understand the impact of DM annihilation on
the CMB at relatively high redshift, i.e. roughly for $z  \gtrsim
100$. 
 At lower redshift, enhanced DM annihilation in non-linear structures might be
responsible for additional effects that we will study separately in the next
section. Since the two regimes have a rather different impact on the CMB
spectra, it is legitimate to split the discussion in this way. DM annihilation
effects on the CMB at high redshift have been thoroughly investigated by Galli
et al. \cite{Galli2009,Galli2011,Finkbeiner2011}. In this section, we will
only update previous results, before exploring new models including halo effects in the next section.

For simplicity, we first assume in subsections \ref{effects_xe},
\ref{annihilation_cl}, \ref{ann_fit} that the annihilation parameter $p_{\rm
ann}$ is independent of redshift, as in \cite{Galli2009,Galli2011}. We will
relax this assumption in subsection~\ref{subsec:variation}.

\subsection{Annihilation effects on $x_e$ and $T_{\rm M}$\label{effects_xe}}

In figure \ref{p_ann}, we show the evolution of $x_e(z)$ and $T_{\rm M}(z)$ computed
with either {\sc recfast} or {\sc hyrec} for four values of the annihilation
parameter.
\FIGURE{
\includegraphics[scale=0.6]{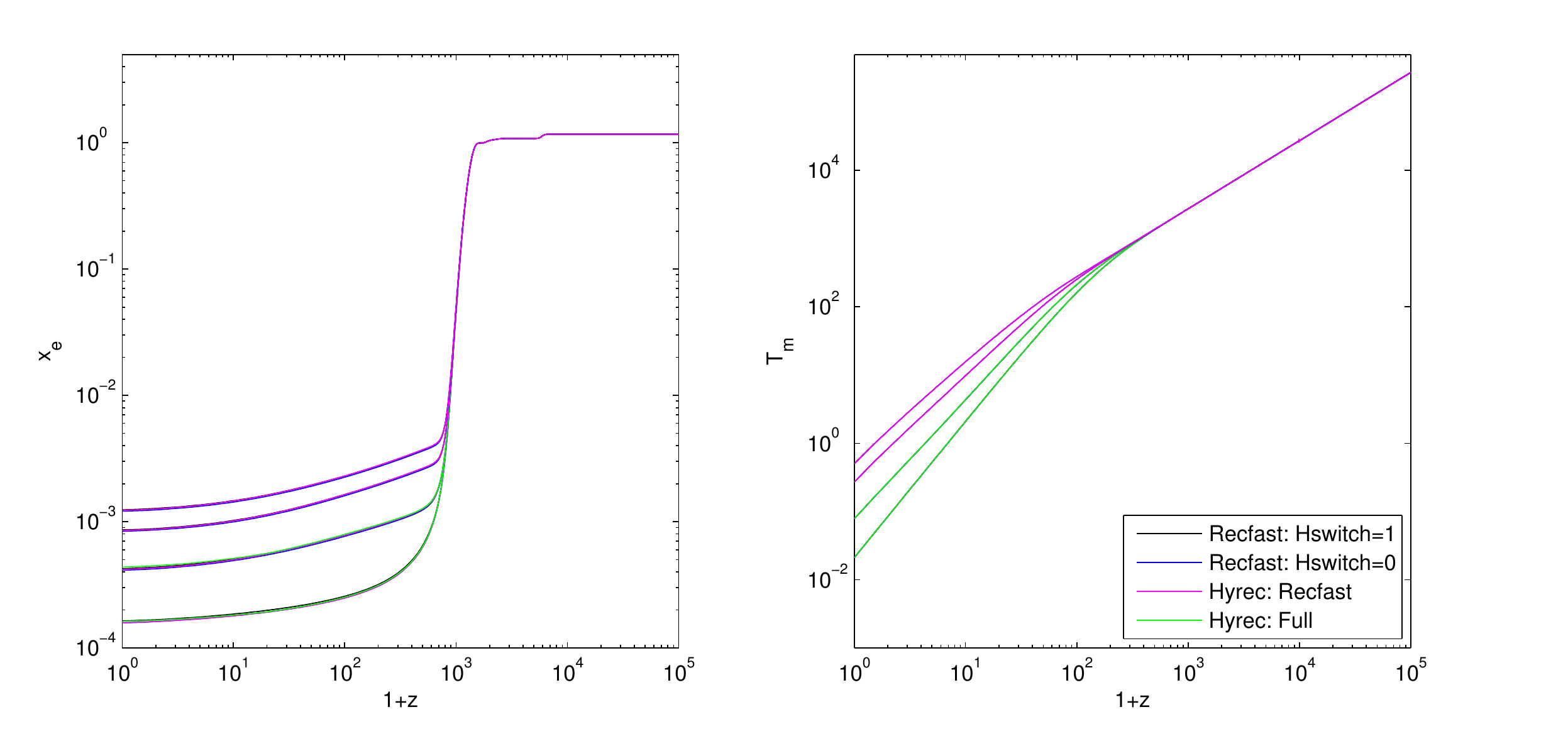}
\caption{Free electron fraction and matter temperature as a function of the redshift with, from bottom to top, $p_{\rm ann}=0, 10^{-6}, 5\cdotp 10^{-6}$ or $10^{-5}$ m$^3$s$^{-1}$kg$^{-1}$. For each value of $p_{\rm ann}$, we used either {\sc recfast} or {\sc hyrec}, and  two different options for each of the two codes; the four results agree to better than a few percent, and the difference would be indistinguishable on the plots.
\label{p_ann}}}
We tested {\sc recfast} and {\sc hyrec} in two modes: for {\sc recfast}, with
or without taking into account the hydrogen physics effects described in
\cite{Scott:2009sz} (using the switch {\tt Hswitch}), and for  {\sc hyrec},
using the mode {\tt RECFAST} (mimicking a simplified version of {\sc recfast})
and {\tt FULL} (including a state-of-the art description of an effective
multi-level hydrogen atom as well as radiative transfer near the Lyman
lines). 
The {\tt FULL} mode uses interpolation tables
requiring ${T_{\rm M}}<{T_r}$. This is the case at all times provided that the
annihilation parameter does not exceed $p_{\rm ann}\leq 3\cdotp 10^{-6}$
m$^3$s$^{-1}$kg$^{-1}$. In order to test {\sc hyrec}/{\tt FULL} above
this value, we removed the condition ${T_{\rm M}}<{T_r}$ from the
code, letting it extrapolate from the table. For all used values of
$p_{\rm ann}$, $T_{\rm M}$ never exceeds $T_r$ by a large fraction and
the extrapolation is therefore accurate.

In the results presented in figure \ref{p_ann}, we assumed a $\Lambda$CDM model
without reionization. The first two small steps seen on the electron fraction
curve correspond to the two helium recombinations, and bring the ratio
$x_e={n_e}/{n_H}$ down to one. The third and biggest step accounts for hydrogen
recombination. As expected, the energy injected by DM annihilation inhibits
recombination, and the free electron fraction freezes out at a larger value.
Moreover, the matter temperature decreases more slowly after photon decoupling due to
energy injection in the gas resulting from DM annihilation.

For each value of $p_{\rm ann}$, the difference between the four algorithms is
extremely smalll\footnote{It would be even smaller using the fudge factor values
  of version 1.5.2 of {\sc recfast}, that was released after the
  submission of this work.}. We checked that  the shifts induced in the CMB power spectra
are well below the sensitivity level of current CMB data sets, and lead to the
same observational bounds on $p_{\rm ann}$. This means that the four approaches
can be used indifferently in the rest of this analysis. Whenever we could, we
sticked to {\sc recfast} with {\tt Hswitch} on, in order to speed up the
computation. We will mention below that for some models, we had to use instead
{\sc hyrec} with the {\tt RECFAST}  or {\tt FULL} mode, found to be the more stable
numerically. In these cases, the increase in computing time in the full
parameter extraction process was less than a factor of two.

\subsection{Effects on the CMB Power spectrum}
\label{annihilation_cl}

We could expect the effect of DM annihilation to be degenerate with
that of reionization, since both mechanisms increase the ionization
fraction after photon decoupling, and therefore the optical depth to
last scattering $\tau(z_{\rm dec})$. Indeed, a high ionization
fraction at $z<z_{\rm dec}$ implies that more photons interact along
the line of sight, which tends to damp temperature and polarization
anisotropies on sub-Hubble scale, and to regenerate extra polarization
around the Hubble scale at the time of re-scattering. 

In figure~\ref{ann_cl_rel}, we compare the effect of varying $p_{\rm ann}$ with that of changing the redshift of reionization, under the usual simplifying assumption of a single reionization step, such that $x_e(z)$ follows a hyperbolic tangent centered on $z_{\rm reio}$. The two effects turn out to be rather different for reasons that are easy to understand. 

\FIGURE{
\includegraphics[scale=0.75]{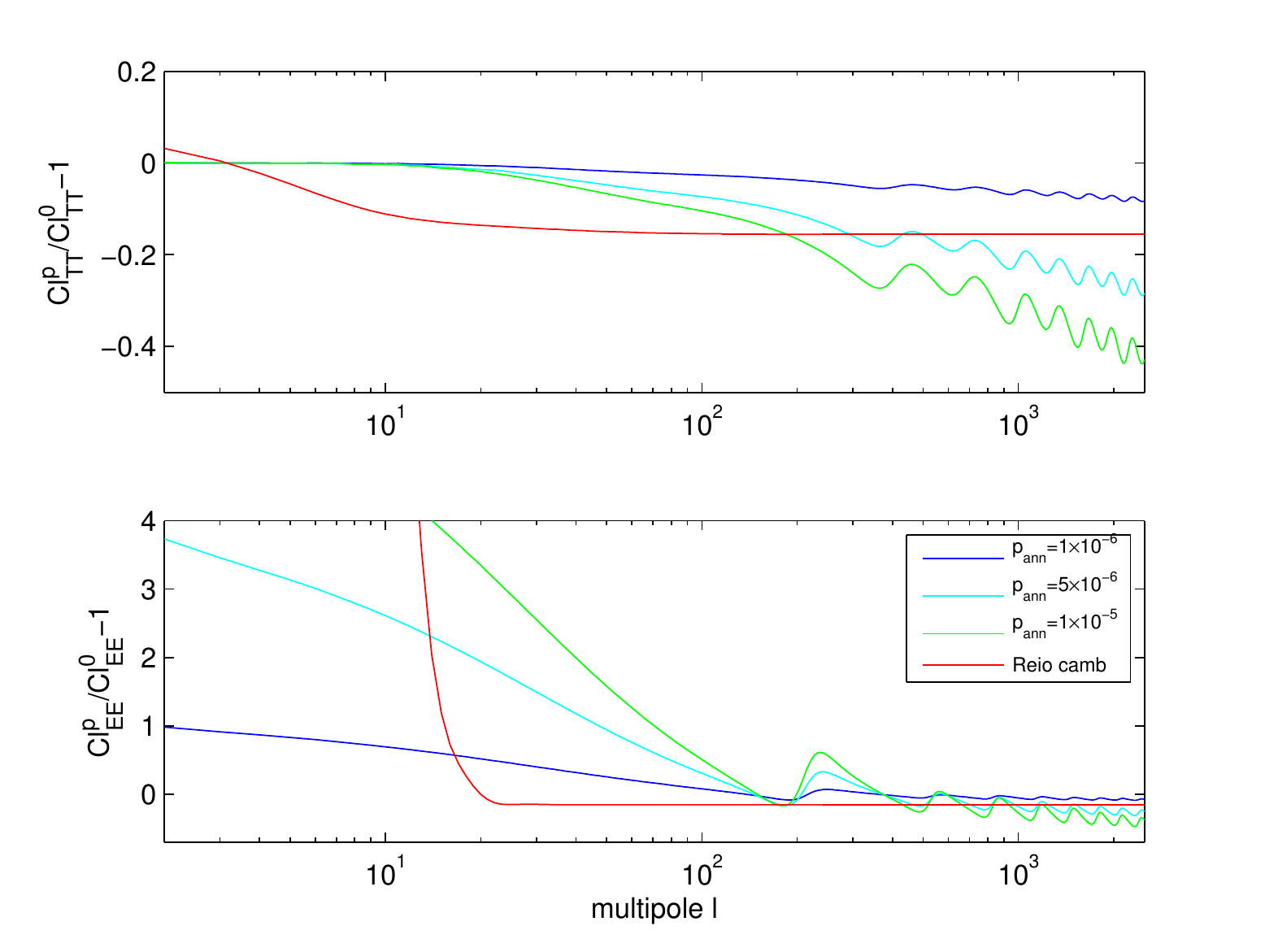}
\caption{Temperature and polarization power spectra for several models with DM annihilation or reionization, rescaled by a reference model with none of them. The curves with oscillatory patterns correspond to  different values of $p_{\rm ann}$ (expressed in the key in m$^3$s$^{-1}$kg$^{-1}$) and no reionization. The last curve was obtained with $p_{\rm ann}=0$ and with reionization at $z_{\rm reio}=11$.
\label{ann_cl_rel}}}

First, the annihilation effect is already present around $z=z_{\rm dec}$, and
results in a small delay in the decoupling time (defined as the maximum of the
visibility function $-\tau'e^{-\tau}$). Hence, the sound horizon at decoupling
has the time to grow, while the diffusion damping scale has sufficient time to
reach larger scales. The increased sound horizon results in peaks visible
under larger angles or smaller $l$'s: this shifting of the peak explains the
oscillatory patterns clearly visible in figure~\ref{ann_cl_rel}. The increased
diffusion damping scale enhances Silk damping at large $l's$, leading to the
negative high-$l$ slope in figure~\ref{ann_cl_rel}.

Second, DM annihilation increases the ionization fraction and the optical depth
at all redshifts in the range $0<z<z_{\rm dec}$. This means that some power is
removed from the temperature and polarization spectrum on all scales, with a
maximum suppression for $l>200$, corresponding to modes being always inside the
Hubble radius in the range  $0<z<z_{\rm dec}$. In the temperature spectrum,
multipoles with $l<200$ are less and less affected when $l$ decreases. In the
polarization spectrum, the rescattering of the photons generate extra
polarization on all scales in the range $2<l<200$ corresponding to the
variation of the Hubble scale between decoupling and today.

In contrast, reionization enhances $x_e(z)$ only at small redshift, $z \leq 10$
in our example. It does not affect recombination and does not shift the peaks:
the reionization curve in figure~\ref{ann_cl_rel} has no oscillatory patterns.
Power is maximally suppressed for all scales being inside the Hubble radius at
reionization, i.e. all multipoles $l>20$. The regeneration of power in the
polarization spectrum is limited to $l<20$ for the same reason (but is very
strong, since reionization enhances $x_e(z)$ much more than DM annihilation).

Hence, DM annihilation effects are clearly not degenerate with reionization
effects. In order to check that the impact of $p_{\rm ann}$ cannot be mimicked by other parameters in the $\Lambda$CDM model, we should however run a parameter extraction code and marginalize the posterior distribution of $p_{\rm ann}$ over other cosmological parameters.

\subsection{Analysis with WMAP and SPT data\label{ann_fit}}

We compared to observations a model described by the six free parameters of the
vanilla $\Lambda$CDM model, the annihilation parameter $p_{\rm ann}$, and the
effective neutrino number $N_{\rm eff}$, accounting \textit{e.g.} for extra
relativistic degrees of freedom. Since the South Pole Telescope (SPT)
collaboration reported an intriguingly high best-fit value of $N_{\rm eff}$
\cite{Keisler:2011aw}, we wish to check whether $p_{\rm ann}$ and $N_{\rm eff}$
are correlated in some way, such that the effect of one parameter could be
confused with that of the other. A priori, this is not impossible, because both
parameters impact the amplitude of the high-$l$ damping tail of the temperature
spectrum, relatively to the amplitude of the first acoustic peaks.

We compared this model to WMAP 7-year data \cite{Komatsu:2010fb} and SPT data
\cite{Keisler:2011aw}, using the code {\sc monte python} \cite{montepython},
based on Monte Carlo Markhov Chains and on the Metropolis-Hastings algorithm
(like {\sc CosmoMC} \cite{Lewis:2002ah}, but {\sc monte python} is interfaced
with {\sc class} instead of {\sc camb} \cite{Lewis:1999bs}, written in {\sc
python}, and has extra functionalities; this software will soon be released
publicly). On top of the cosmological parameters, we vary three nuisance
parameters related to the foreground contamination of the SPT data and
constrained by gaussian priors, following strictly the recommendations and the
software released by the SPT collaboration. All results on cosmological
parameters are marginalized over these three nuisance parameters.  We took flat
priors on all parameters and just imposed $p_{\rm ann}>0$. 

Our results, summarized in the first column of Table~\ref{table1} and in the
triangle plot of figure~\ref{fig_p_ann}, are in excellent agreement with those
of the SPT collaboration for the first seven parameters (last column of Table 3
in ref.~\cite{Keisler:2011aw}). For DM annihilation, we obtain a bound
\begin{equation}
p_{\rm ann}< 0.89 \times 10^{-6}{\rm m}^3/{\rm s}/{\rm kg} \qquad {\rm (WMAP7+SPT, ~95\% C.L)}. 
\end{equation}
We observe no correlation between $p_{\rm ann}$ and any other parameter in the
analysis (in particular, we checked that there is no correlation at all with
$N_{\rm eff}$). The marginalized posterior probability for $p_{\rm ann}$ is
displayed on the left plot of figure~\ref{fig_p_ann}, and shows no evidence for
DM annihilation in current data. Our bound is stronger than the most recent
one, presented in ref.~\cite{Galli2011}, 
\begin{equation}
p_{\rm ann}< 2.09 \times 10^{-27}{\rm cm}^3/{\rm s}/[{\rm GeV}/c^2] =
1.17 \times 10^{-6}{\rm m}^3/{\rm s}/{\rm kg} \qquad {\rm (WMAP7+ACT, ~95\% C.L)},
\end{equation}
due to the inclusion of the SPT dataset. It is also stronger than that from ref.~ \cite{Hutsi:2011vx}. We refer the reader to ref.~\cite{Galli2011} for a discussion of derived limits on the DM annihilation cross-section, given various ansatz for the mass and released energy fraction $f$.

\TABLE{
\begin{tabular}{|c|c|c|c|c|c|}
\hline
annihilation &\multirow{2}{*}{neglected}&\multirow{2}{*}{neglected}&\multirow{2}{*}{yes}&\multirow{2}{*}{yes}&\multirow{2}{*}{yes}\\
in halos:&&&&&\\
\hline
reionization &\multirow{2}{*}{yes}&\multirow{2}{*}{yes}&\multirow{2}{*}{neglected}&\multirow{2}{*}{yes}&\multirow{2}{*}{yes}\\
from stars:&&&&&\\
\hline
\multirow{2}{*}{data:}&\multirow{2}{*}{CMB}&\multirow{2}{*}{CMB}&\multirow{2}{*}{CMB}&CMB +&CMB +\\
&&&&Gunn-Pet.&$T_{M}$ prior\\
\hline
section:&3.3&3.4&4.4&\ref{gunn_peterson}&\ref{tigm}\\
\hline
\hline
&&&&&\\
100 $\omega_b$& $2.282^{+0.055}_{-0.055}$&$2.281^{+0.054}_{-0.057}$&$2.267^{+0.052}_{-0.052}$&$2.267_{-0.053}^{+ 0.052}$&$2.280_{-0.055}^{+ 0.055}$\\
&&&&&\\
$\omega_{cdm}$&$0.125_{-0.013}^{+0.011}$&$0.125^{+0.011}_{-0.013}$&$0.126^{+0.011}_{-0.013}$&$0.126_{-0.013}^{+0.011}$&$0.126_{-0.013}^{+0.011}$\\
&&&&&\\
$n_s$&$ 0.987_{-0.020}^{+0.020}$&$0.987^{+0.020}_{-0.020}$&$0.980^{+0.019}_{-0.019}$&$0.980_{-0.018}^{+0.019}$&$0.985_{-0.021}^{+0.019}$\\
&&&&&\\
$10^{9}A_s$&$2.39_{-0.12}^{+0.11}$&$2.39^{+0.11}_{-0.12}$&$2.44^{+0.11}_{-0.13}$&$2.42_{-0.12}^{+0.11}$&$2.38_{-0.12}^{+0.11}$\\
&&&&&\\
$h$&$0.753_{-0.042}^{+0.038}$&$0.753^{+0.038}_{-0.042}$&$0.750^{+0.036}_{-0.040}$&$0.751_{-0.039}^{+0.037}$&$0.760_{-0.043}^{+0.039}$\\
&&&&&\\
$N_{\rm eff}$&$3.84_{-0.67}^{+0.60}$&$3.85^{+0.66}_{-0.60}$&$3.89^{+0.59}_{-0.64}$&$3.88_{-0.65}^{+0.59}$&$3.96_{-0.67}^{+0.58}$\\ 
&&&&&\\
$z_{\rm reio}$&$10.9_{-1.4}^{+1.3}$&$10.9^{+1.3}_{-1.4}$&-&$6.58_{-0.09}^{+0.10}$&$12.2_{-1.6}^{+1.6}$\\
&&&&&\\
$\frac{10^{6}p_{\rm ann}}{{\rm m}^3/{\rm s}/{\rm kg}}$&$<0.89$&$<0.91$&$<0.78$&$<0.75$&$<0.78$\\
&&&&&\\
$\alpha$&-&flat&-&-&-\\
&&&&&\\
$\frac{f_{\rm h}}{{\rm m}^3/{\rm s}/{\rm kg}}$&-&-&$12600^{+4100}_{-8800}$&$13000_{-8400}^{+3100}$&$<1400$\\
&&&&&\\
$z_{\rm h}$&-&-&$23.4_{-8.4}^{+2.7}$&$20.7_{-5.2}^{+3.7}$&flat\\
&&&&&\\
\hline
%&&&&&\\
$[-2 \ln {\cal L}]_{\rm min}$ & 3752.7 $\times$ 2 & 3752.7 $\times$ 2 & 3753.1 $\times$ 2 & 3753.2 $\times$ 2& 3752.7 $\times$ 2\\
%&&&&&\\
\hline
\end{tabular}
\caption{Mean and edges of the 68\% Minimum Credible Interval (MCI) for the cosmological parameters of the five models that we compared to WMAP7 and SPT data. We don't show results for the nuisance parameters associated to SPT data, that have been marginalized over. The second model differs from the first one by the inclusion of a $z$-dependent annihilation function parametrized by $\alpha$. All parameters have been assigned top-hat priors, and never reach prior edges except $p_{\rm ann}$ (limited to positive values), $\alpha$ (limited to the range $-0.2<\alpha<0$) and $z_{\rm h}$ (on which we imposed a prior $20 \leq z_{\rm h} \leq 30$ only in the last column).
For $p_{\rm ann}$ (and $f_{\rm h}$ in the last column),  we indicate the 95\% Confidence Level (C.L.) upper bound.
\label{table1}}}

\FIGURE{
\includegraphics[scale=0.35]{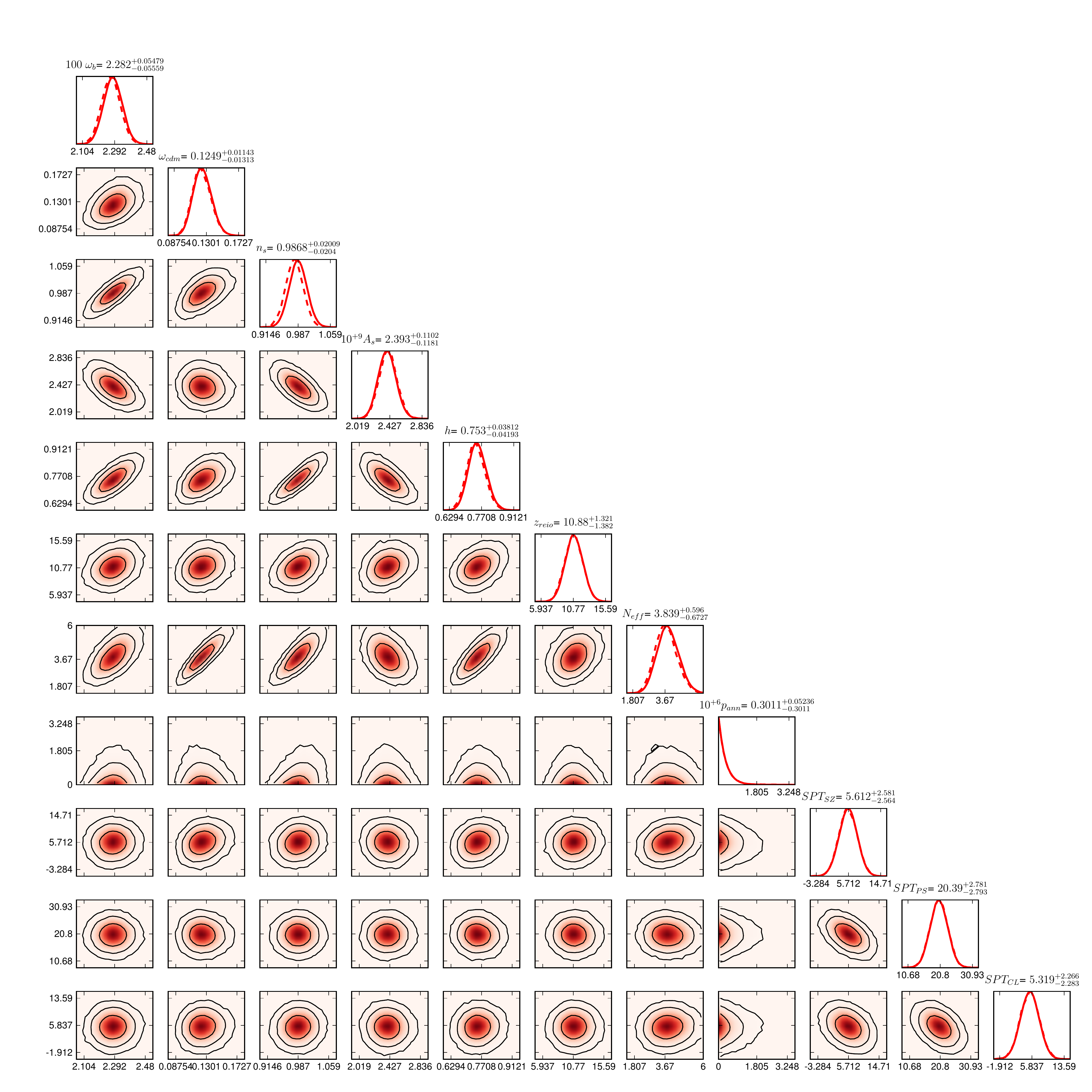}
\caption{One and two-dimensional marginalized posterior probabilities for the free parameters of a $\Lambda$CDM model with a free effective neutrino number $N_{\rm eff}$ and time-independent annihilation parameter $p_{\rm ann}$, compared to WMAP7 + SPT data. For the two-dimensional posterior, we show the contours corresponding to the 68.3\%, 95.4\% and 99.7\% credible regions. The last three parameters are nuisance parameters accounting for foregrounds contributions to the SPT data set.
\label{fig_p_ann}}}

%%%%%%%%%%%%%%%%%%%%%%%%%%%%%%%%%%%%%%%%%%%%%
\subsection{Redshift dependent annihilation parameter\label{subsec:variation}}

In any realistic model,  the fraction of energy absorbed by the overall gas is
a function of the redshift $f=f(z)$, as shown in Figure 4 of Slatyer et
al.~\cite{Slatyer2009} for several examples. The shape of $f(z)$ depends on the
DM annihilation channel(s). The impact of the redshift-dependence of $f(z)$ on
current/future CMB constraints on DM annihilation has been questioned with
different methods in various papers
\cite{Hutsi:2011vx,Galli2011,Finkbeiner2011}. Here we will check this issue
with yet another approach, and confirm the results of other references showing
that taking this dependence into account is of very little relevance.

All examples for $f(z)$ shown in  \cite{Slatyer2009} have strong similarities:
$f(z)$ is always a smooth step-like function, with plateaux at $z>2500$ and
$z<30$. In view of performing a model-independent comparison to the data, it is
tempting to approximate $f(z)$ with a family of simple analytic functions,
capturing the essential behavior of $f(z)$ in all cases. The CMB is marginally
affected by the behavior of $f(z)$ at low $z$ even in the case of a constant
$p_{\rm ann}$, and even more if $f(z)$ decreases; moreover, the effect of DM
annihilation at low $z$ is superseded by that of reionization. Hence, a given
approximation scheme doesn't need to be accurate at low $z$, but should capture
the essential behavior for $z>100$. Figure 4 in  \cite{Slatyer2009} suggests
that  $f(z)$ could be chosen to be constant at $z>2500$, to decrease like a
parabola in log-log space for $30<z<2500$, and to remain again constant at
$z<30$, as displayed in figure~\ref{z-dep}.
\FIGURE{
\includegraphics[scale=0.6]{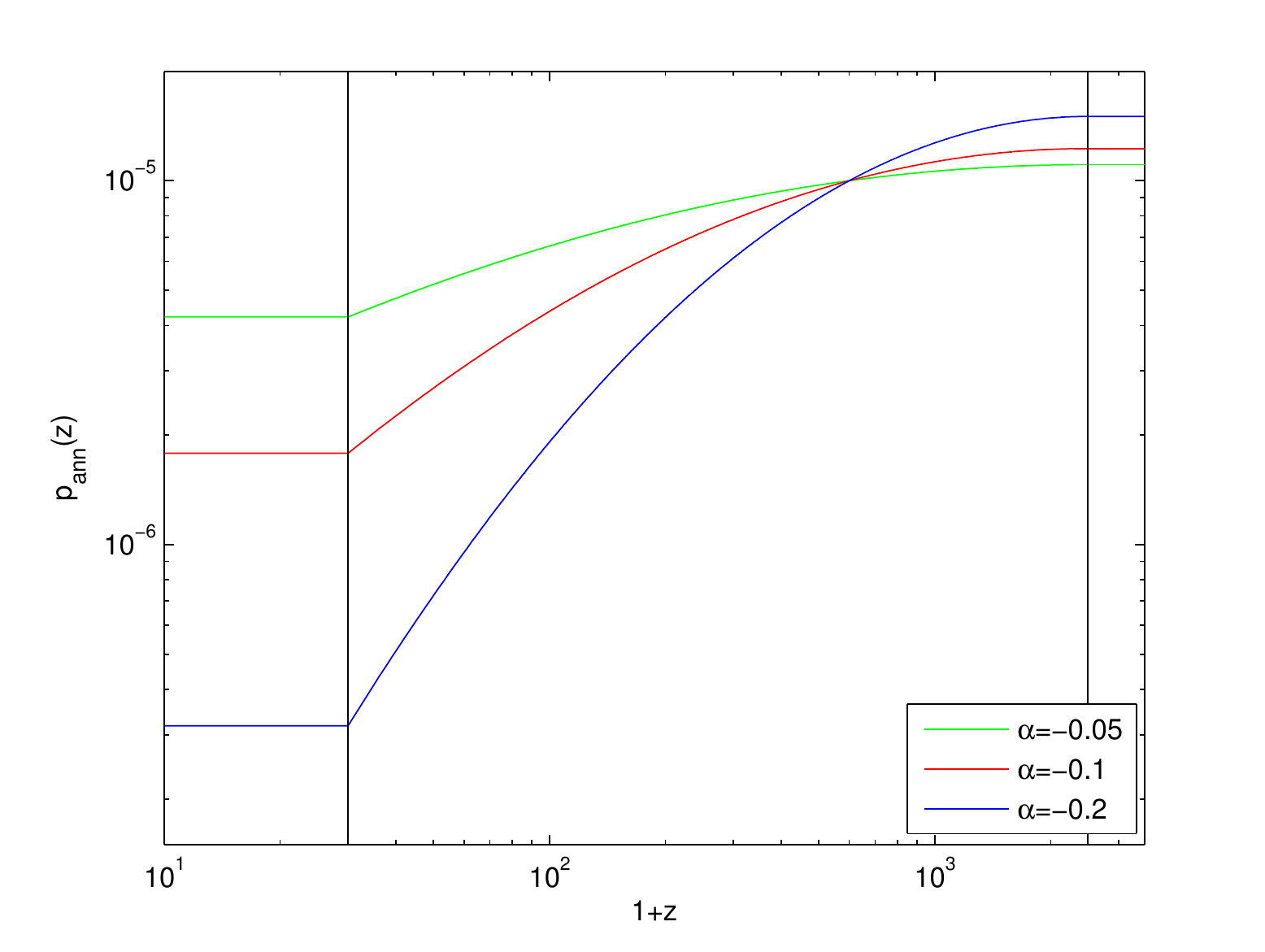}
\caption{Redshift dependent $p_{\rm ann}(z)$, approximated as a two-parameter family of functions as described in the text, with 
$\bar{p}_{\rm ann} \equiv p_{\rm ann}(z=600)=1\cdotp 10^{-5}$ m$^3$s$^{-1}$kg$^{-1}$ and 
$\alpha=-0.05, -0.1, -0.2$.
\label{z-dep}}}

The work of \cite{Finkbeiner2011} analyzed the amount of information that one
can extract from current and future CMB data on $f(z)$ or $p_{\rm ann}(z)$ (we
recall that these two functions are simply related to each other by a
time-independent factor, as long as we assume that the annihilation cross section does not vary with temperature). A model-independent analysis, based on the expansion
of $p_{\rm ann}(z)$ in principal components, reveals that  the CMB is mainly
sensitive to the first principal component, peaking around $z=600$, and at the
next order to the second principal component, accounting for the redshift
variation of $f(z)$ around this same value. 

The goal of this section is to check these results with a simpler approach than
a full principal component analysis. We will stick to the simple approximation
for $p_{\rm ann}(z)$ suggested above, involving two plateaus and one parabola.
This family of functions has two free parameters, one amplitude and one
curvature. We can choose to define the amplitude parameter $\bar{p}_{\rm ann}$
at $z=600$:
\begin{equation}
p_{\rm ann}(z)=\left\{\begin{aligned} &\bar{p}_{\rm ann}\exp\left[-\alpha\log^2\left(\frac{601}{2501}\right)\right] & \text{ for } z>2500,
\\ &\bar{p}_{\rm ann}\exp\left[\alpha\left(\log^2\left(\frac{1+z}{2501}\right)-\log^2\left(\frac{601}{2501}\right)\right)\right] & \text{ for } 30<z<2500,
\\ &\bar{p}_{\rm ann}\exp\left[\alpha\left(\log^2\left(\frac{31}{2501}\right)-\log^2\left(\frac{601}{2501}\right)\right)\right] & \text{ for } z<30,
\end{aligned}\right.
\end{equation}
with $\bar{p}_{\rm ann} \equiv f(z=600) {\langle \sigma v\rangle}/{m_{\rm DM}}$ and $\alpha<0$. With respect to the previous section, we now have a new dimensionless parameter $\alpha$, that expresses the redshift dependence of $p_{\rm ann}(z)$.
The question is whether this new parameter can be detected with current data: if not, the analysis of the previous section captures all the information that we can extract, with $p_{\rm ann}$ standing for the value of the annihilation parameter near $z\sim 600$.

We show in figure~\ref{xe(z)} the evolution of $x_e$ and $T_{\rm M}$ for fixed $\bar{p}_{\rm ann}$ and several values of $\alpha$. As long as $\alpha$ remains small in absolute value ($|\alpha| \ll 1$), its impact is mainly on the slope of $x_e(z)$ in the region $z \ll z_{\rm dec}$. We expect this slope to be difficult to probe experimentally, since the CMB is mainly sensitive to the optical depth, which is an integrated quantity over redshift.

When the redshift dependance of $p_{\rm ann}(z)$ increases with a fixed normalization at $z=600$, the annihilation rate at high redshift increases. We expect to reach such large values that the decoupling time is not just slightly affected by DM annihilation, but radically postponed to a later time, because the massive energy injection from DM annihilation forbids hydrogen recombination. This happens for $\alpha<-3$, as illustrated in figure~\ref{xe(z)}. In this regime, the sound horizon at recombination is dramatically increased, and the CMB data will enforce a similar increase in the angular diameter distance to last scattering, in order to keep the same peak scale in multipole space. This will generate a correlation between $\alpha$ and parameters such as the Hubble rate. However, the scale of the acoustic peaks and of Silk damping react differently to such a transformation, so we expect that $\alpha$ cannot be pushed to arbitrary negative values. This ``extreme'' regime could not be reached in the previous subsection: as long as we assumed a constant $p_{\rm ann}$, observational bounds on $p_{\rm ann}$ prevented the annihilation rate to be too high around $z\sim 1000$. When comparing this model with CMB data, we first imposed no prior on $\alpha$ (apart from $\alpha<0$). We obtained a bound $\alpha>-5.3$ (95\% C.L.) and some non-trivial correlation between very negative values of $\alpha$ and other parameters.

\FIGURE{
\includegraphics[scale=0.6]{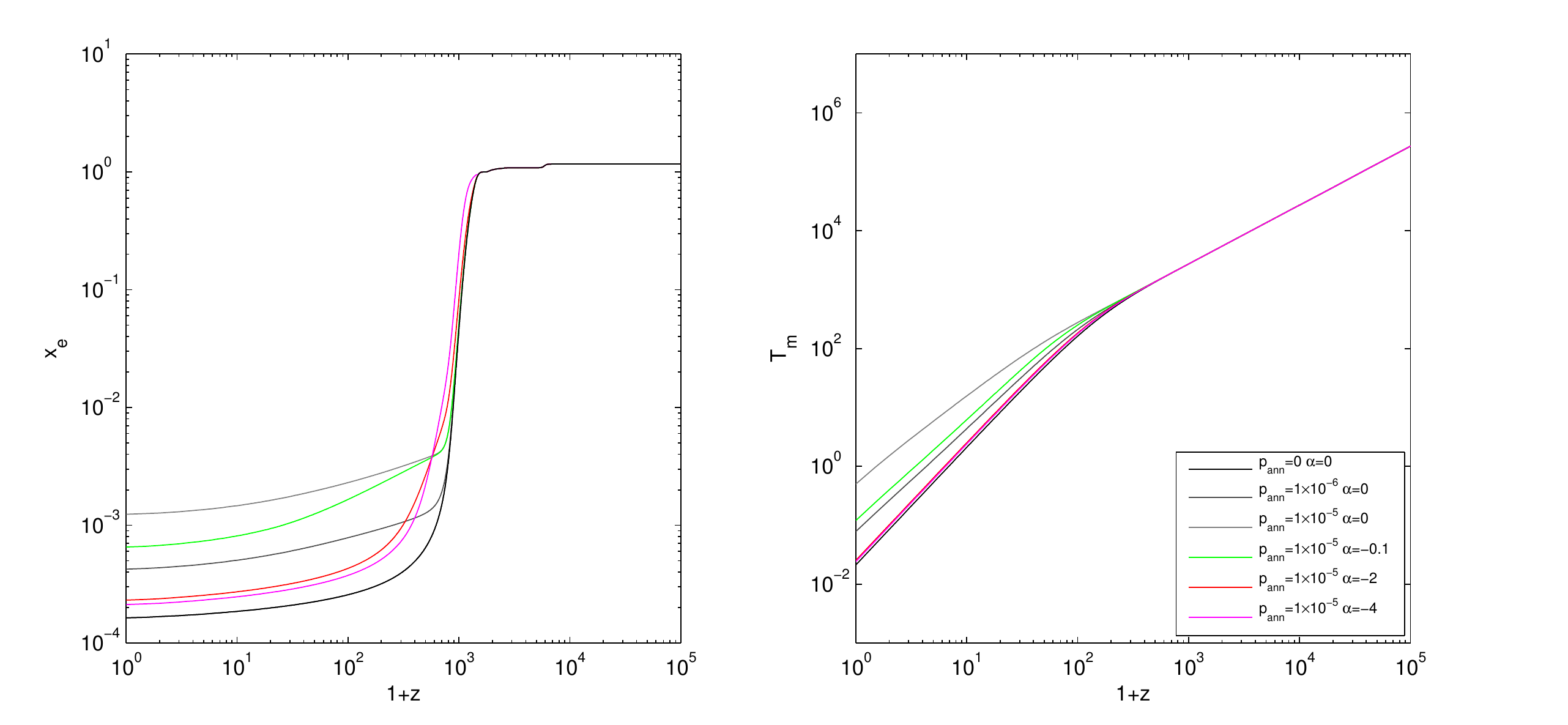}
\caption{Free electron fraction $x_e$ and matter temperature $T_{\rm M}$ as a function of redshift for a constant $p_{\rm ann}=0,\ 1\cdotp 10^{-6}$ and $1\cdotp 10^{-5}$ m$^3$s$^{-1}$kg$^{-1}$(black and gray curves) and a redshift dependent $p_{\rm ann}(z)$ with $\bar{p}_{\rm ann}=1\cdotp 10^{-5}$ m$^3$s$^{-1}$kg$^{-1}$ and $\alpha=-0.1, -2$ and $4$, using {\sc recfast} and assuming no reionization.
\label{xe(z)}}}

However, this region in parameter space should not be taken seriously, because
the realistic examples provided in~\cite{Slatyer2009} correspond to values of
$|\alpha|$ at most of the order of 0.1 or 0.2. We performed a more
``realistic'' run with a top-hat prior $-0.2<\alpha<0$. The results are
summarized in the second column of Table~\ref{table1}. The data still gives no
indication in favor of DM annihilation. The posterior probability of $\alpha$
is flat throughout the prior range, and the bounds on other parameters are
essentially unchanged with respect to the run with a constant annihilation
parameter, i.e. with $\alpha=0$. Even the two-dimensional probability contours
in the space $(\bar{p}_{\rm ann}, \alpha)$ show no significant correlation
between these parameters. 

These results are fully consistent with those of  \cite{Finkbeiner2011},
showing that the first principal component peaks near $z\sim600$. They also
prove  that current data is not sensitive to the second principal component,
unless it has an unreasonably large amplitude like in the run with no prior on
$\alpha$. We could have defined our parameter $\bar{p}_{\rm ann}$ at a
different redshift: in that case, we would have expected to find a correlation
between $\bar{p}_{\rm ann}$ and $\alpha$. The maximum of the first principal
component can be seen as the ``decorrelation redshift'' between $\bar{p}_{\rm
ann}$ and $\alpha$.

In conclusion of this section, it appears that the moderate variation of $f(z)$ (or equivalently $p_{\rm ann}(z)$) in the range $40<z<1000$ suggested by the realistic examples of \cite{Slatyer2009} is far from being detectable with WMAP7+SPT data. Ref.~\cite{Galli2011,Hutsi:2011vx} reached a similar conclusion by comparing bounds on $p_{\rm ann}$ for
some particular cases out of the possible $f(z)$ functions presented in \cite{Slatyer2009}. In the next section, it will be legitimate to neglect any variation of $p_{\rm ann}(z)$ at least until the redshift of halo formation.

%%%%%%%%%%%%%%%%%%%%%%%%%%%%%%%%%%%%%%%%%%%%
\section{Annihilation in Dark Matter halos and Reionization\label{sec:halos}}

Until now we considered that Dark Matter is uniformly distributed in the universe. It is well-known that  structure formation generates a concentration of DM in halos that is likely to enhance DM annihilation. This issue has been widely discussed in the context of dark matter indirect detection in cosmic rays. It has also been pointed out that enhanced DM annihilation could be relevant for the reionization of the universe, and therefore for CMB physics~\cite{Natarajan:2008pk,Belikov2009,Cirelli2009,Natarajan2010,Hutsi:2011vx,Natarajan2012}. In this section, we wish to propose a systematic investigation of such effects, based on a generic parameterization of DM annihilation in halos, and a full parameter extraction from CMB data.

\subsection{Energy density release in DM halos}
The energy released by Dark Matter annihilation in halos can be written as (see
e.g. \cite{Belikov2009,Natarajan2012})
\begin{equation}
\left.\frac{d E}{dV dt}\right|_{\rm halos}= \rho^2_\chi (z) c^2 p_{\rm ann},
\label{eq_halo1}
\end{equation}
where $\rho^2_\chi$ is the squared dark matter density averaged over space, that exceeds the square of the average dark matter density in presence of non-linear structures. In the halo model, this quantity is given by:
\begin{equation}
\rho^2_\chi (z)=(1+z)^3 \int_{M_{\rm min}}^\infty dM \frac{dn}{dM} \left(\int_0^{r_{200}} dr 4\pi r^2\rho_{\rm h}^2(r)\right).
\label{rho_halo}
\end{equation}
Here $M_{\rm min}$ is the minimal mass of DM halos,  $\frac{dn}{dM}$ the
differential comoving number density of DM halos of mass $M$, $r_{200}$ the
radius of a sphere enclosing a mean density equal to 200 times the background
density, and $\rho_{\rm h}$ the spherical DM halo density profile. The shape of
density profiles is still a subject of controversy. If we consider for instance
a Navarro-Frenk-White (NFW) profile \cite{Navarro1995}, we can express the last
integral as
\begin{equation}
\int_0^{r_{200}} dr 4 \pi r^2 \rho_{\rm h}^2(r)=\frac{M \bar{\rho}(z_{\rm F})}{3} \left(\frac{\Omega_{\rm DM}}{\Omega_{M}}\right)^2 f_{\rm NFW}(c_{\rm h}),
\end{equation}
where $z_{\rm F}$ is the redshift of halo formation, $\bar{\rho}(z_{\rm F})=200 \rho_{\rm c} \Omega_{\rm M} (1+z_{\rm F})^3$ the average matter density within a radius $r_{200}$, and $f_{\rm NFW}$ a function of the so-called halo concentration parameter $c_{\rm h}$. We recall that the critical density $\rho_{\rm c}$ and density fraction parameters $\Omega_i$ are defined today. In order to get an analytic approximation to $dn/dM$, one can use the Press-Schechter formalism \cite{Press1974}, leading to
\begin{equation}
\frac{dn}{dM}(M,z)=\frac{\rho_{\rm M}}{M} \frac{d \ln \sigma^{-1}}{dM} f(\sigma),
\end{equation}
where $\rho_{\rm M}=\rho_{\rm c} \Omega_{\rm M}$ is the average matter density today, and $f(\sigma)$ the differential mass function. The variance of the linear density field $\sigma (M,z)$ is given as usual by
\begin{equation}
\sigma^2 (M, z)=\int_0^\infty P(k,z) W^2(k,M) k^2 dk,
\end{equation}
with $P(k,z)$ the linear power spectrum at redshift $z$, and $W(k,M)$ the window function. Assuming that the collapse of the high density regions can be described by a spherical model, one can use a top-hat filter for $W$ \cite{Lukic2007}.
For the differential mass function, we could rely on the original function of Press and Schechter
\begin{equation}
f_{PS}(\sigma)= \sqrt{\frac{2}{\pi}} \frac{\delta_{sc}}{\sigma} \exp\left(-\frac{\delta_{sc}^2}{2 \sigma^2} \right)
\end{equation}
with $\delta_{sc}=1.28$,
or the more accurate function proposed by Seth and Thormen \cite{Sheth1999}.
We could compute these terms exactly within the Boltzmann code, but the CMB spectra are not highly sensitive to the details of the halo model: they can only provide constraints on integrated quantities. Hence, it is irrelevant to search for high accuracy in this context. Instead, it would be very useful for the purpose of fitting CMB data to derive a simple, approximate parametric form for the energy injection function. To start with, we can use the fact that in a universe dominated by matter (i.e. any time between decoupling and $z \sim 1$), the redshift dependence of the variance $\sigma$ is somewhat trivial:
\begin{equation}
\sigma (M, z)= \sigma (M,1) \frac{2}{1+z}.
\end{equation}
\\ If we recollect all terms, we get the following contribution to the energy rate due to Dark Matter halos
\begin{align}
\left.\frac{d E}{dV dt}\right|_{\rm halos}=&\rho_{\rm c}^2\,  \Omega_{\rm DM}^2 \, c^2 \, p_{\rm ann}(z) \,  (1+z)^3 \, \frac{200}{3} (1+z_{\rm F})^3 f_{\rm NFW}(c_{\rm h}) \nonumber \\
&\quad \times \int_{M_{\rm min}}^\infty dM \bigg\lbrace \frac{ d\ln \sigma^{-1} (M,1)}{dM}
\frac{2}{\sqrt{\pi}}
\frac{\delta_{sc}\, (1+z)}{2 \sqrt{2} \, \sigma (M,1)}
 \exp\left(-\frac{\delta_{sc}^2 \, (1+z)^2}{8 \, \sigma^2 (M,1)} \right)\bigg\rbrace,
 \label{eq_halo2}
\end{align}
where we used the original Press-Schechter differential mass function for simplicity.
The redshift-dependent integral simplifies with the change of variable  $u=\frac{\delta_{sc}(1+z)}{2\sqrt{2}\sigma(M,1)}$. If we define $u_{\rm min}(z)=\frac{\delta_{sc} (1+z)}{2\sqrt{2}\sigma(M_{\rm min},1)}$, it reduces to
\begin{equation}
\int_{u_{\rm min}(z)}^\infty  du \frac{2}{\sqrt{\pi}} \exp \left( -u^2 \right)= {\rm erfc}(u_{\rm min}(z))~,
\end{equation}
where $\textrm{erfc}(x)$ is the complementary error function. It is suppressed at high redshift, before halo formation, i.e. as long as $u_{\rm min}(z)\gg1$. At low redshift, we do not expect the function $p_{\rm ann}(z)$ to vary significantly, as can be seen in figure 4 of  \cite{Slatyer2009} for several examples. Hence, we can replace it by a nearly constant value $p_{\rm ann}(0)$. In this case, the energy rate from annihilation in  halos can be expressed as a function of only two parameters (beyond $\rho_{\rm c} \Omega_{\rm DM}$),
\begin{equation}
\left.\frac{d E}{dV dt}\right|_{\rm halos}\simeq \rho_{\rm c}^2 \Omega_{\rm DM}^2 c^2 \, (1+z)^3 \,f_{\rm h}\,\,
\textrm{erfc}\left(\frac{1+z}{1+z_{\rm h}}\right)~,
\end{equation}
where $z_{\rm h}$ is the characteristic redshift at which halos start to contribute\footnote{In fact, the functions $\textrm{erfc}(x)$ starts to raise at $x \leq 2$, so halos contribute below $z\leq 2z_{\rm h}$.},\begin{equation}
z_{\rm h} \equiv \frac{2 \sqrt{2}}{\delta_{sc}}    \sigma(M_{\rm min},1)-1~,
\end{equation}
and $f_{\rm h}$ is a parameter related to the formation redshift and concentration of halos, and to the DM annihilation function at low redshift,
\begin{equation}
f_{\rm h} \equiv \frac{200}{3}(1+z_{\rm F})^3 \, f_{\rm NFW}(c_{\rm h}) \, p_{\rm ann}(0)~.
\end{equation}
The parameter $f_{\rm h}$ shares the same units as $p_{\rm ann}$.
In principle, $z_{\rm h}$ and the ratio $f_{\rm h}/p_{\rm ann}(0)$ should be inferred from a fit of the Press-Schechter formalism to detailed simulations of structure formation. However, there is no full consensus yet on the dynamics of halo formation and on halo density profiles. Moreover, these parameters should have a strong dependence on cosmological parameters, and also on  the matter power spectrum at large $k$, which is poorly constrained by observations. Hence we will treat $z_{\rm h}$ and $f_{\rm h}$ as free parameters in our analysis. 

\subsection{Beyond the on-the-spot approximation\label{beyond}}

Sticking to the on-the-spot approximation, we could express the net ionization rate per interval of redshift $I_{X_i}+I_{X_\alpha}$ using eqs.~(\ref{ionization},\ref{excitation}). The equation of evolution for $x_e$ and $T_{\rm M}$ would receive extra contributions
\begin{align}
\left.\frac{d x_e}{dz}\right|_{\rm ann} &= - \frac{1}{(1+z) H(z) n_H(z)} \frac{d E}{dV dt}(z)
 \left[ \frac{1-x_e(z)}{3} \left(\frac{1}{E_i}+\frac{1-C(z)}{E_\alpha}\right)\right],
\label{onthespot}\\
\left. \frac{dT_{\rm M}}{dz} \right|_{\rm ann} &= - \frac{1}{(1+z) H(z) n_H(z)} \frac{d E}{dV dt}(z)\left[\frac{2}{3 k_b}\frac{1+2x_e(z)}{3}\frac{1}{1+f_{He}+x_e(z)}  \right],
\label{Tmonthespot}
\end{align}
with $\frac{d E}{dV dt}(z)$ given by the sum of the smooth density and halo density contributions. However, this approximation becomes inaccurate at low redshift, as we shall see below. Well after recombination, the energy injection responsible for ionization and heating at a given redshift $z$ comes from the decay products of DM annihilation at all redshifts $z' \geq z$, taking into account the fact that particle energies are redshifted between $z'$ and $z$, and that a fraction of the particles created at $z'$ interact along the line-of-sight and do not play a role at $z$. Moreover, in general, the interaction cross-section between $z'$ and $z$ depends on the energy of each particle. Hence, the ionization rate obeys to a rather complicated equation involving two integrals: one over $dz'$, and one over the energy of the particles created at $z'$, and interacting with the plasma at intermediate redshift. 

However, ref.~\cite{Belikov2009} found that at low redshift, most of the ionization and heating is caused by photons produced by the inverse Compton scattering (ICS) of charged particles resulting from DM annihilation over CMB photons. It was shown by the authors of~\cite{Natarajan2010} that taking mainly this process into account leads to the simplified expression:
\begin{align}
\left.\frac{d x_e}{dz}\right|_{\rm ann} &= - \frac{c \sigma_T \, \gamma(z)}{(1+z) H(z)}  \int_z^\infty \frac{dz'}{(1+z')H(z')}
\left(\frac{1+z}{1+z'}\right)^3 e^{\kappa(z,z')}
\frac{d E}{dV dt}(z'),
\label{fullxe}
\end{align}
where $e^{\kappa(z,z')}$ is an absorption factor: it represents the fraction of photons produced around $z'$ by ICS that already interacted with the inter-galactic medium and deposited their energy before $z$. Hence $\kappa$ can be approximated as
\begin{align}
\kappa(z,z') &= c \sigma_T \int_z^{z'} \frac{-dz''}{(1+z'')H(z'')} n_H(z'')~,
\end{align}
not to be confused with the optical depth of CMB photons, featuring an extra factor $x_e(z'')$.
Finally, $\gamma(z)$ is a short-cut notation for
\begin{align}
\gamma(z) \equiv  \left[ \frac{1-x_e(z)}{3} \left(\frac{1}{E_i}+\frac{1-C(z)}{E_\alpha}\right)\right].
\end{align}
The integral in $\kappa(z,z')$ can be performed analytically:
\begin{align}
\kappa(z,z')
=  \frac{2}{3} c \sigma_T  \frac{n_H(0)}{H_0\sqrt{\Omega_M}}  \left[(1+z)^{3/2} - (1+z')^{3/2}\right] = \frac{2}{3} c \sigma_T \left[\frac{n_H(z)}{H(z)}-\frac{n_H(z')}{H(z')}\right]~.
\end{align}
Let us define $\alpha \equiv c \sigma_T  \frac{n_H(0)}{H_0\sqrt{\Omega_M}}$. We notice that
\begin{align}
\int_z^\infty dz' \, \alpha \, \sqrt{1+z'} \, e^{\kappa(z,z')} = 1~.
\end{align}
Hence, the function $\Delta(z,z') \equiv \alpha \, \sqrt{1+z'} \, e^{\kappa(z,z')}$ peaking in $z'=z$ can be approximated with the Dirac function $\delta(z-z')$ in the limit in which it decreases with $z'$ much faster than any other function in the integrand of equation (\ref{fullxe}). Writing (\ref{fullxe}) as
\begin{align}
\left.\frac{d x_e}{dz}\right|_{\rm ann} &= - \frac{\gamma(z)}{(1+z)H(z)} \int_z^\infty dz'\,\,\Delta(z,z')\,\,n_H(z')^{-1}\left(\frac{1+z}{1+z'}\right)^3
\frac{d E}{dV dt}(z')\,\,,
\end{align}
we see that in the approximation mentioned above, one recovers exactly the on-the-spot expression of eq.~(\ref{onthespot}). But in the general case, we have to deal with the full integral.
This is mathematically equivalent to keeping expression (\ref{onthespot}), with the on-the-spot energy rate replaced by an effective one, defined as
\begin{align}
\left.\frac{d E}{dV dt}\right|_{\rm eff}\!\!\!(z)
&\equiv \int_z^\infty dz' \,\, \Delta(z,z') \,\, \frac{n_H(z)}{n_H(z')}
\left(\frac{1+z}{1+z'}\right)^3 
\frac{d E}{dV dt}(z')\nonumber\\
&= \int_z^\infty dz'
\Delta(z,z')
\left(\frac{1+z}{1+z'}\right)^{6}
\frac{d E}{dV dt}(z')~.
\label{xefull}
\end{align}
Similarly, this effective energy rate should be used in equation (\ref{Tmonthespot}) to get the correct temperature evolution. 
\\
\FIGURE{
\includegraphics[width=7cm,angle=-90]{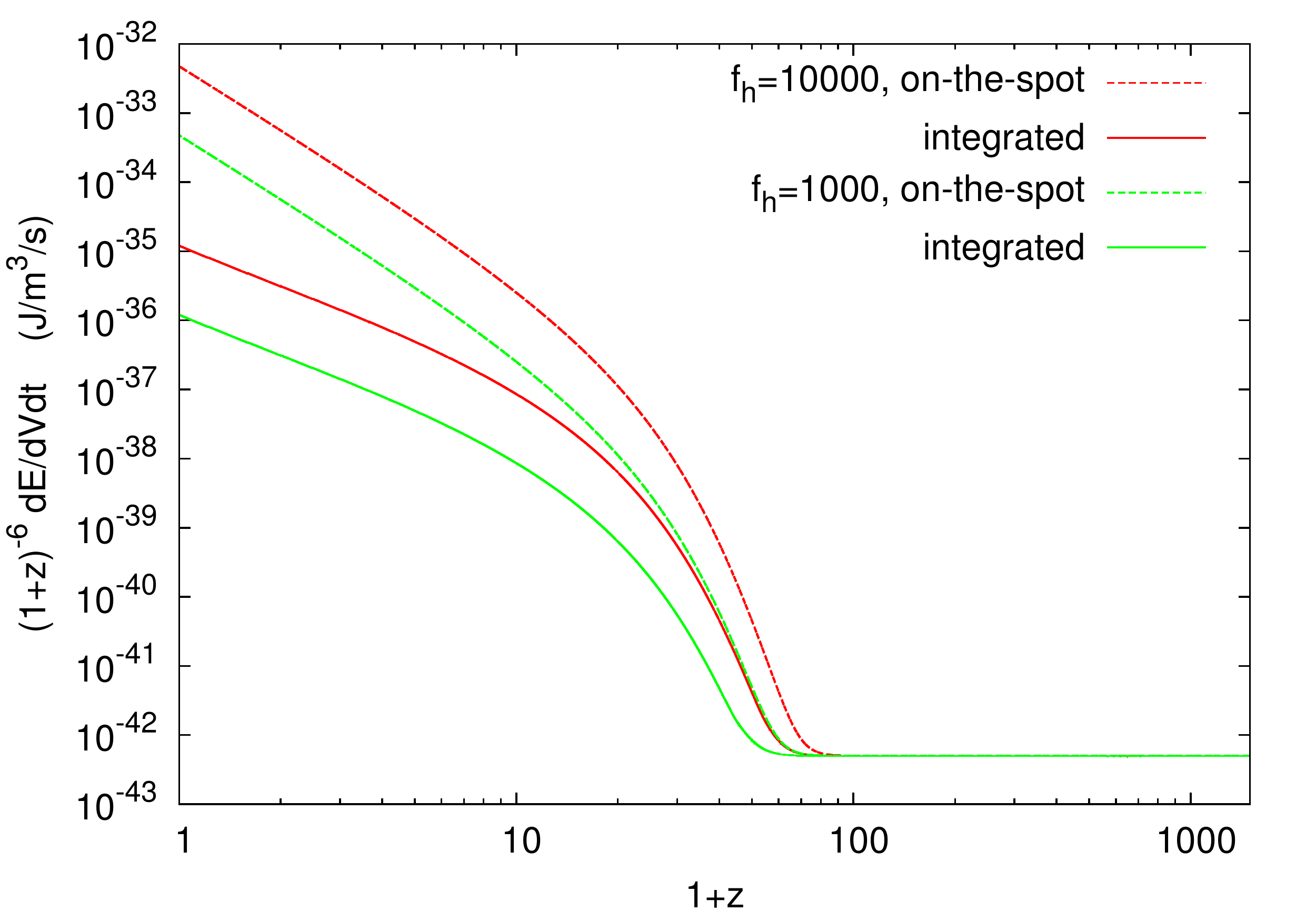}
\caption{\label{fig_dxe} Effective energy injection rate rescaled by $(1+z)^6$, computed with or without the on-the-spot approximation, for either $f_{\rm h}=10000$ or $1000$. Other annihilation parameters are fixed to $p_{\rm ann}=10^{-6}$ and $z_{\rm h}=20$.}}

\subsection{Effects on $x_e$, $T_{\rm M}$ and the CMB spectra}

In figure~\ref{halos_xe}, we compare the effect of DM annihilation in halos with that of the default reionization model implemented in {\sc class} and {\sc camb}, based on a hyperbolic tangent centered in $z_{\rm reio}$. The effect of DM annihilation on the ionization fraction is found to be very similar, except that it induces a slower reionization. The parameter $z_{\rm h}$ controls the onset of reionization from halos, while $f_{\rm h}$ controls its amplitude. For large enough values of $f_{\rm h}$, DM annihilation in halos can entirely reionize the universe before the current epoch, as shown previously in \cite{Belikov2009,Natarajan2010,Natarajan2012}. With the default reionization model, the ionization fraction $x_e$ is larger than one at low redshift, because Helium reionization is also taken into account. In our model for DM annihilation in halos, we neglect helium reionization for simplicity, so that $x_e$ is smaller or equal to one by definition.
\FIGURE{
\includegraphics[scale=0.4,angle=-90]{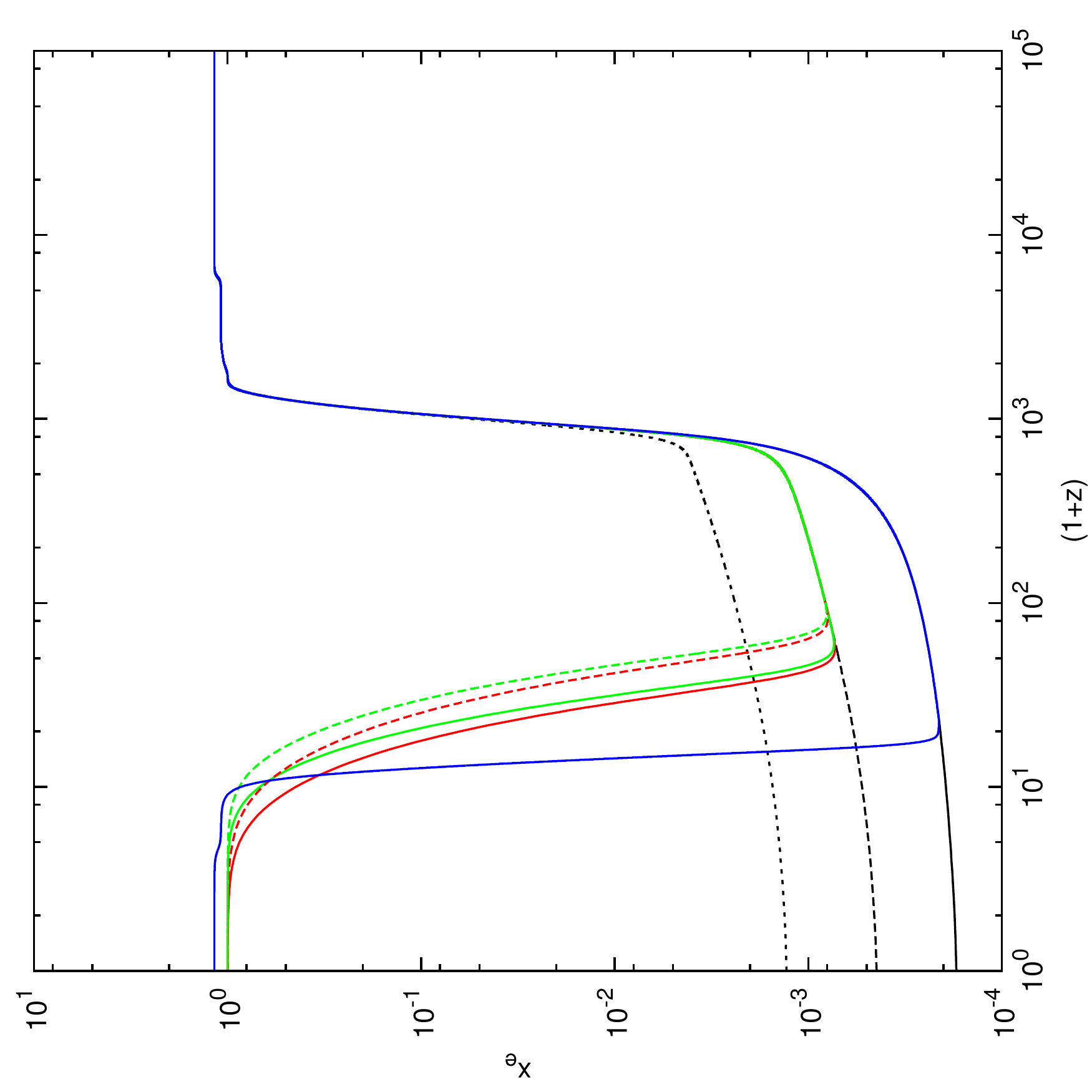}
\includegraphics[scale=0.4,angle=-90]{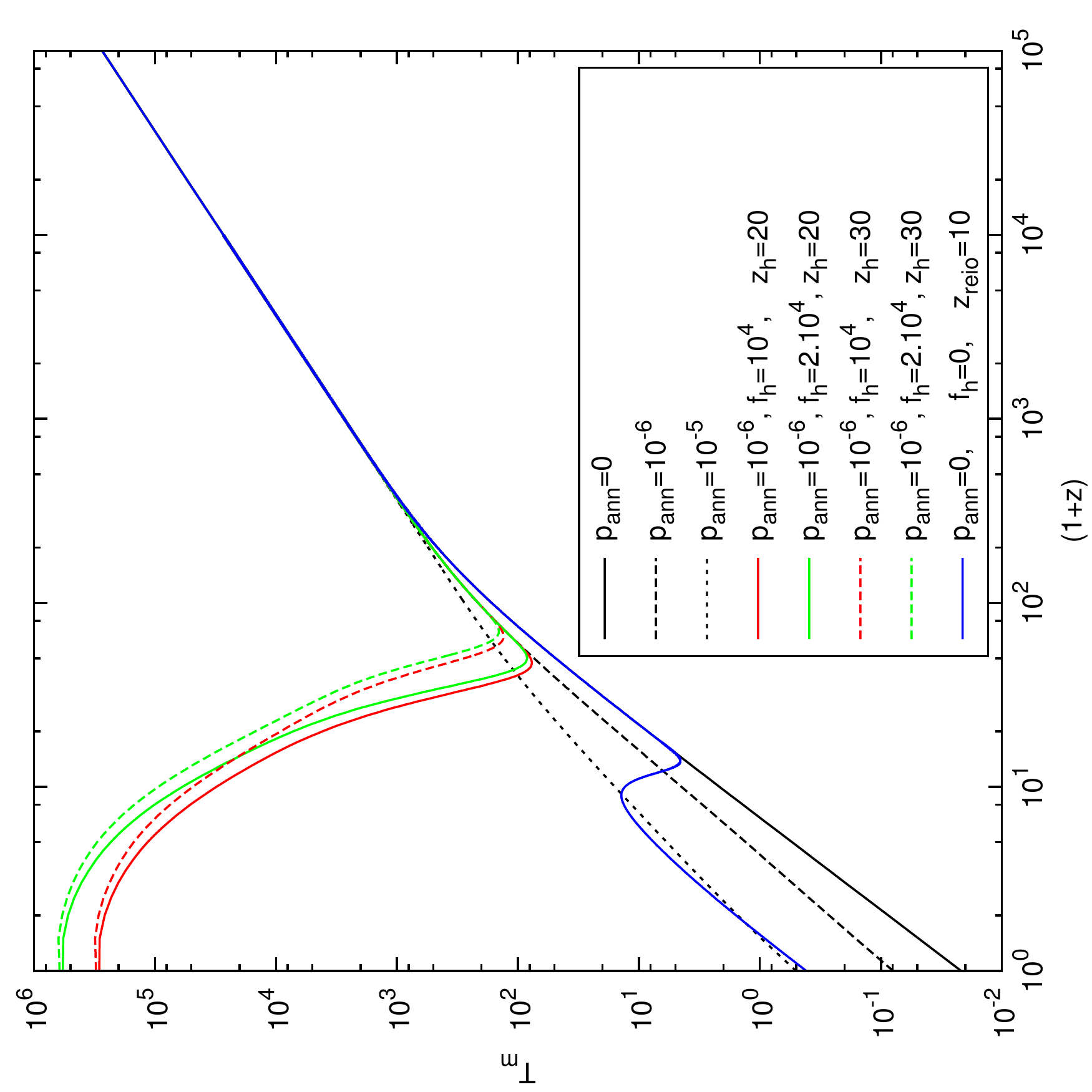}
\caption{Free electron fraction and matter temperature for $p_{\rm ann}=0, 10^{-6}$ and $10^{-5}$  m$^3$s$^{-1}$kg$^{-1}$ (from bottom to top) and different values of $f_{\rm h}$ and $z_{\rm h}$, compared to the usual results for $p_{\rm ann}=0$ and a single-step model for reionization from stars. All curves were obtained using {\sc hyrec} in mode {\tt RECFAST}.
\label{halos_xe}}}

When including the effect of DM annihilation in halos, we work with the {\tt RECFAST} mode of {\sc hyrec}. Indeed,  with {\sc recfast}, we experienced numerical instability issues: the free electron fraction explodes and oscillates very rapidly already for small value of our parameters $f_{\rm h}$ and $z_{\rm h}$. With {\sc hyrec} in {\tt FULL} modes, the only problem is that for large values of $z_{\rm h}$ and $f_{\rm h}$, the ratio $T_{\rm M}/T_r$ may exceed one, falling outside the range of one interpolation tables. The {\tt RECFAST} mode of {\sc hyrec} is always well behaved.

The right plot in figure~\ref{halos_xe} shows that the matter temperature increases a lot due DM annihilation in halos. Note also that for extreme values of the temperature $T_M>2\times10^{4}$~K, using {\sc RECFAST}'s case-B recombination coefficient becomes inaccurate~\cite{1991A&A...251..680P}. We will see anyway in section~\ref{tigm} that such large values are in contradiction with constraints on the temperature of the inter-galactic medium at $z\leq4$, as inferred from Lyman-$\alpha$ observations: this will provide an addition constraint on the DM annihilation rate.

The signature of DM annihilation on the primary CMB anisotropy spectrum is found to be very similar to that of reionization. In addition to the peak shifting and damping due to a non-zero $p_{\rm ann}$ parameter, the halo effect controlled mainly by $f_{\rm h}$ leads to an overall suppression of temperature/polarization power for $l>30$, and an enhancement of polarization  for $l<30$. We can anticipate that the CMB alone can hardly discriminate between the contribution of reionization from stars and from halos, since the CMB spectra probe mainly the optical depth, i.e. the integral of $x_e$ over time. However, the fact that DM induces a slow reionization process starting at high redshift\footnote{In the CMB analysis of the next subsections, $z_{\rm h}$ is found in the range from 20 to 30, implying that halos start contributing between 40 and 60, well before star formation.} implies that the step-like suppression of temperature and the low-$l$ polarization bump are smoother and wider than with the default reionization model. To illustrate this, we compare in figure~\ref{bestfit_reio} the low-$l$ polarization spectrum for two models with the same optical depth. Accurate CMB polarization data limited only by cosmic variance on large angular scale may probe such a difference.
\FIGURE{
\includegraphics[width=7cm,angle=-90]{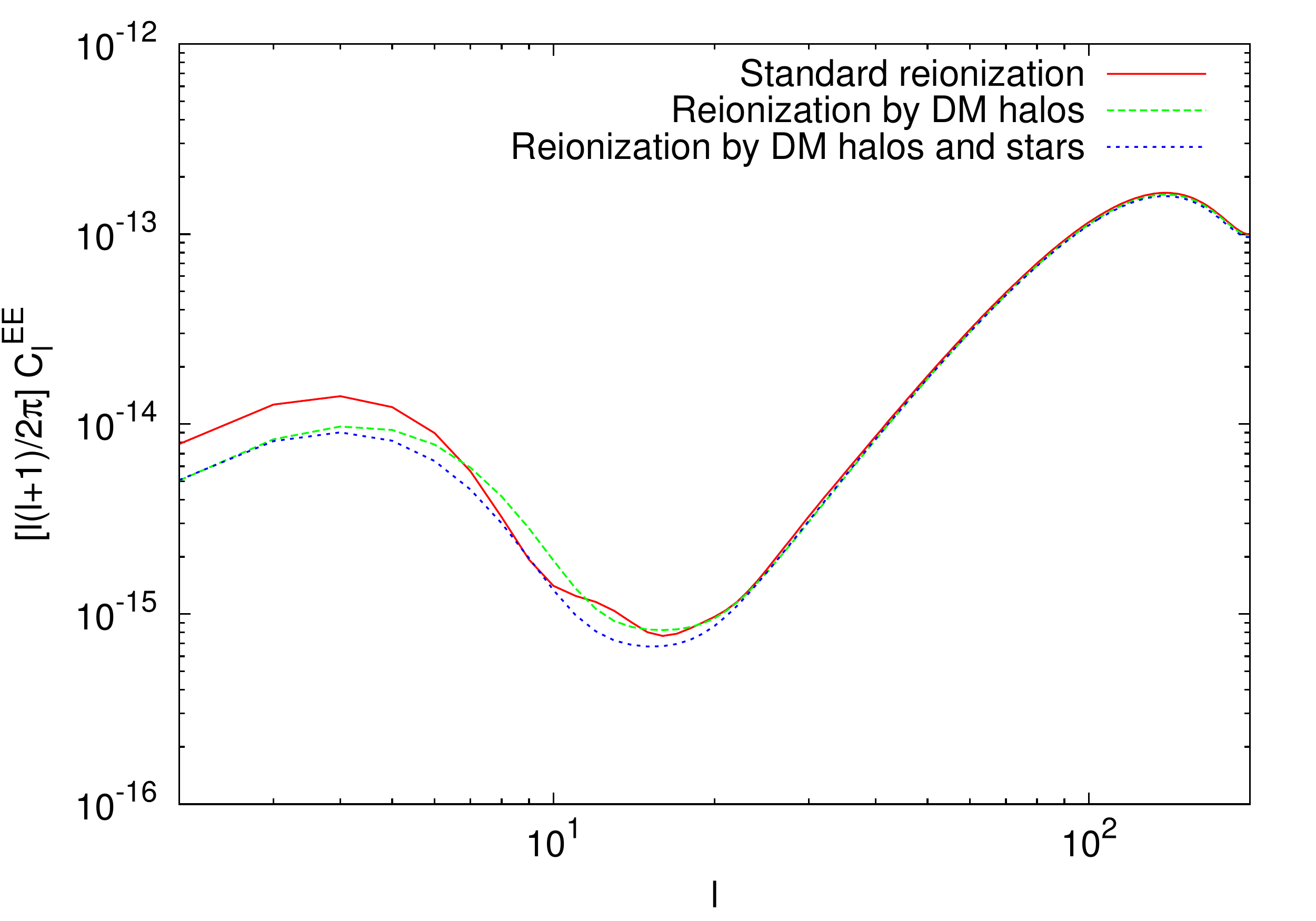}
\caption{Low-$l$ polarization spectrum for the three best-fitting models assuming reionization from stars (with the usual single step parameterization), from DM annihilation in  halos, or from both with an additional Gunn-Peterson prior.
\label{bestfit_reio}}}

\subsection{Can Dark Matter annihilation alone explain reionization?\label{reio_halo}}

We wish to check whether WMAP7 and SPT data are compatible with the assumption that the reionization of the universe can be explained entirely by DM annihilation in halos, as suggested in \cite{Belikov2009,Natarajan2010,Natarajan2012}. It is rather obvious that the free parameters $z_{\rm h}$ and $f_{\rm h}$ of our model can be adjusted in such way that the reionization optical depth is be compatible with the WMAP7 best-fitting value. However, we have seen that DM annihilation can only induce slow reionization starting at high redshift, and induce a wider step (resp. bump) in the low-$l$ temperature (resp. polarization) spectrum. A priori, this may lead to a   value of the maximum likelihood significantly lower for the annihilation model. In addition, an analysis with free $z_{\rm h}$ and $f_{\rm h}$ could lead to preferred values in strong contradiction with expectations from structure formation and halo models.

The results of our {\sc montepython} run with WMAP and SPT data are summarized in the third column of Table~\ref{table1} and in the
triangle plot of figure~\ref{halos_monte}. The new free parameters $z_{\rm h}$ and
$f_{\rm h}$ are not degenerate with other parameters, so the credible interval for
the usual $\Lambda$CDM parameters and $N_{\rm eff}$ are unchanged with respect
to the standard model without annihilation. There is instead a significant
correlation between $z_{\rm h}$ and $f_{\rm h}$: if halos form very late, a very large
amplitude parameter $f_{\rm h}$ is needed in order to get the same optical depth. The
effective chi square $\chi^2_{\rm eff} \equiv -2 \ln {\cal L}$ is higher for DM
reionization than single-step star reionization, but only by 0.8, showing
that the data shows no strong preference for one model against the other. 

The characteristic redshift $z_{\rm h}$ is found in the range $12<z_{\rm h}<40$ (95\%C. L.). This parameter has a strongly non-gaussian posterior probability, with a mean value of 23, but a best-fit value of 19. The shape of the $\textrm{erfc}(x)$ function is such that the halo contribution starts raising around $z\sim 2 z_{\rm h}$.  Values of $z_{\rm h}$ close to 20 imply a raise in the range 40-50,  which is plausible from the point of view of structure formation. 

To see whether the required value of halo concentration is sensible, we need to make an assumption about the DM annihilation amplitude, since the data is sensitive to $f_{\rm h}$, i.e. to the product $f_{\rm NFW}(c_{\rm h}) \times p_{\rm ann}(0)$.  
If we first assume a value of $p_{\rm ann}$ at $z \sim 600$ saturating our CMB bound, 
$p_{\rm ann}(600) \sim 9\times 10^{-7}$,  
we expect that at low redshift this parameter will fall to approximately
$p_{\rm ann}(0) = \sim 2\times10^{-7}$. 
Then the best-fitting value $f_{\rm h} \sim 12600\,{\rm m}^3/{\rm s}/{\rm kg}$ requires 
\begin{equation}
f_{\rm NFW}(c_{\rm h}) \sim \frac{3 f_{\rm h}}{200 (1+z_{\rm F})^3 p_{\rm ann}(0)} \sim 4200~,
\end{equation}
where we also assumed $z_{\rm F} \sim 60$. The quantity $f_{\rm NFW}(c_{\rm h})$ is poorly constrained, but models of halo formation suggest an order of magnitude ranging from $10^3$ to $10^5$. Hence the ``reionization from DM annihilation'' model points towards a reasonable value of the concentration parameter. If $f_{\rm NFW}(c_{\rm h})$ is of the order of $4\times10^3$, then constraints on DM annihilation from the smooth background and from halos are comparable. If $f_{\rm NFW}(c_{\rm h})$ is of the order of $10^3$ (resp. $10^4$ or $10^5$), then constraints from annihilation in the smooth background (resp. in halos) are stronger. Indeed, the bound on $p_{\rm ann}$ coming from annihilation from halos can be found using the relation
\begin{equation}
p_{\rm ann}(z=600) = 3.3 \times 10^{-11} \, \frac{p_{\rm ann}(600)/p_{\rm ann}(0)}{0.5} \frac{61^3}{(1+z_{\rm F})^3} \frac{10^4}{f_{\rm NFW}(c_{\rm h})} f_{\rm h}~\label{rel_pann_f}.
\end{equation}
Taking 
%$f_{\rm h}^{\rm max}=24500\,{\rm m}^3/{\rm s}/{\rm kg}$ 
$f_{\rm h}^{\rm max}=25600\,{\rm m}^3/{\rm s}/{\rm kg}$ (95\% C.L.) from our analysis, this implies
\begin{equation}
p_{\rm ann}(z=600) < 0.84 \times 10^{-6} \left[ \frac{p_{\rm ann}(600)}{0.5 p_{\rm ann}(0)}\frac{61^3}{(1+z_{\rm F})^3} \frac{10^4}{f_{\rm NFW}(c_{\rm h})}\right] {\rm m}^3/{\rm s}/{\rm kg}~~(95\% \textrm{C.L.})
\label{halo_constraint}
\end{equation}

\FIGURE{
\includegraphics[scale=0.22]{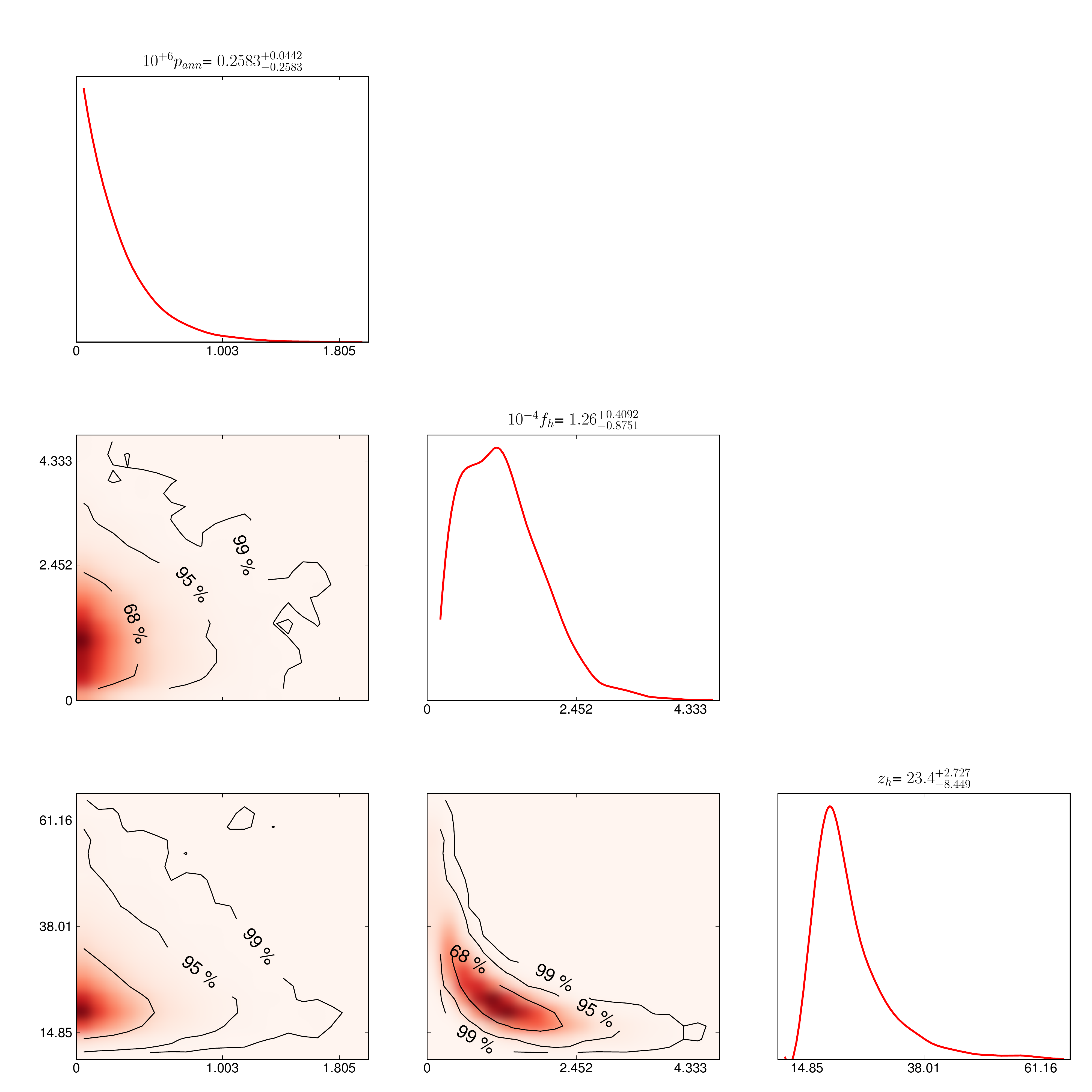}
\caption{Assuming a $\Lambda$CDM model with free electron fraction, dark matter annihilation (background and halos) and no extra reionization from stars, marginalized probability distribution of the annihilation parameters $p_{\rm ann}$, $f_{\rm h}$ and $z_{\rm h}$ given WMAP7 and SPT data.
\label{halos_monte}}}

The low-$l$ polarization spectrum for the best fitting model is shown on figure \ref{bestfit_reio}. Given its distinct shape due to an early and slow reionization process (with respect to star reionization), we expect future small-scale polarized measurements by Planck and other CMB experiments to improve the bound on $f_{\rm h}$.

\subsection{The Gunn-Peterson effect\label{gunn_peterson}}

It was realized in 1965 by Gunn and Peterson \cite{Gunn1965} that the observation of redshifted Lyman-$\alpha$ absorption lines in quasar spectra was a very sensitive probe of the presence of neutral hydrogen along the line of sight, and hence of the ionization fraction of the universe at different redshifts. Since even a small fraction of neutral hydrogen leads to a clear signature, we have some evidence that the universe was almost fully ionized until $z\sim6$, since quasars at such a redshift show a very small level of  Lyman-$\alpha$ absorption. More precisely, according to \cite{Fan2006}, the fraction of neutral hydrogen $x_{\rm HI}$ has to satisfy:
\begin{itemize}
\item for $z\geq6$, $x_{\rm HI}\geq10^{-3}$ ($x_{\rm HI}$ might even be equal to $10^{-1}$), 
\item for $z\leq 5.5$, $x_{\rm HI}\leq 10^{-4}$,
\end{itemize}
(see also \cite{Caruana} for a recent constraint at $z\simeq7$).
Thus there seems to be an abrupt transition between $z=5.5$ and  $z=6$. This raises some tension with the simplest model of single-step reionization from stars, in which the ionization fraction is assumed to evolve roughly like a hyperbolic tangent centered on a reionization redshift $z_{\rm reio}$. The problem is not related so much to the precise shape of the step, but to the fact that any abrupt step should be centered near $z=6$ or 7 to comply with Gunn-Peterson observations, instead of $z_{\rm reio}=10.6 \pm 2.4$ (95\%C.L.) to explain the optical depth $\tau_{\rm reio} = 0.088 \pm 0.015$ measured by WMAP \cite{Komatsu:2010fb}\footnote{To be more precise, the CMB constraint is dominated by the measurement of low-$l$ E-type polarization by WMAP, and depends on the assumed cosmological model}. A single-step reionization at $z\sim 10$ or even 8 would lead to $x_{\rm HI}<10^{-3}$ at $z=6$.

The model of the previous subsection, in which reionization is caused entirely by DM annihilation in halos, also fails to explain  Gunn-Peterson observations for the opposite reason: reionization is then so slow that all allowed models have $x_{\rm HI}>10^{-4}$ at $z=5.5$.

There could be several solutions to this problem:
\begin{itemize}
\item the Gunn-Peterson bounds may be wrong or not correctly interpreted. These bounds are in fact model-dependent and controversial, since they rely on assumptions concerning the density and temperature of the inter-galactic medium, and the ultra-violet background. Observations at $z\sim6$ could be explained with alternative models for the IGM and UV background, instead of incomplete reionization  \cite{Becker:2006qj}. The evidence that the universe is fully ionized below $z\sim5.5$ could also disappear with different assumptions, for instance in the context of inhomogeneous reionization \cite{McGreer:2011dm}.
\item the cosmological model describing our universe may have extra ingredients (not necessarily related to reionization) such that a good fit to WMAP data can be obtained with single-step reionization at $z_{\rm reio} \sim 6$ or 7.
\item
reionization may be caused by different population of stars forming at different redshifts. The single-step model is too naive and should be replaced by a model with at least two steps. The late one should take place around $z\sim6$ or 7  to account for Gunn-Peterson observations. The early one, possibly related to  the generation of massive, metal-free stars \cite{Cen:2002zc,Wyithe:2002qu}, should partially reionize the universe and enhance the optical depth. 
\item 
reionization may be caused both by star and by the decay or annihilation of some particles. The possibility of enhancing reionization with sterile neutrino decay has been proposed by \cite{Hansen:2003yj}. In the case of annihilating DM, ref.~\cite{Natarajan2010} suggested that  DM annihilation in halos may start to slowly reionize the universe. At a redshift close to six, star formation processes take over and quickly ionize the remaining hydrogen atoms.
\end{itemize}
In this subsection, we wish to test the last paradigm. It is a priori not obvious that any model of this type can work, because in order to explain the observed optical depth, DM annihilation may need to be so large that in any case $x_{\rm HI}<10^{-3}$ at $z=6$. Fortunately, we will see that this mixed model nicely complies with Gunn-Peterson and CMB constraints, and points to plausible halo parameter values. 

We added a Gunn-Peterson prior to WMAP7 and SPT data and ran {\sc montepython} again. More precisely, we impose two top-hat priors $10^{-3} \leq x_{\rm HI}(6) \leq 1$ and $0\leq x_{\rm HI}(5.5) \leq 10^{-4}$. We neglect Helium reionization for simplicity. In this approximation, $x_e$ just represents the fraction of ionized hydrogen, and we have $x_{\rm HI} = 1-x_e$. Our Boltzmann code {\sc class} simulates mixed reionization in the following way. For each model, the ionization fraction is first computed down to $z=0$ neglecting reionization from stars, using {\sc hyrec} (in {\tt RECFAST} mode) in order to avoid numerical instability. The effect of stars is then implemented ``by hand'': below some arbitrary redshift $z_{\rm reio}$, $x_e(z)$ is cut and matched continuously to a half-hyperbolic tangent centered on $z_{\rm reio}$, reaching an asymptotic value of one for $z \rightarrow 0$ (see one example of such models in figure~\ref{halos_mixed_xe}). The precise shape of this function is in fact identical to that for ordinary single-step reionization in {\sc class} and {\sc camb}, except that only the side $z\leq z_{\rm reio}$ of the step-like function is used, and that the transition width parameter is decreased to $\delta z = 0.2$ in order to model a very fast reionization process.
\FIGURE{
\includegraphics[width=7cm,angle=-90]{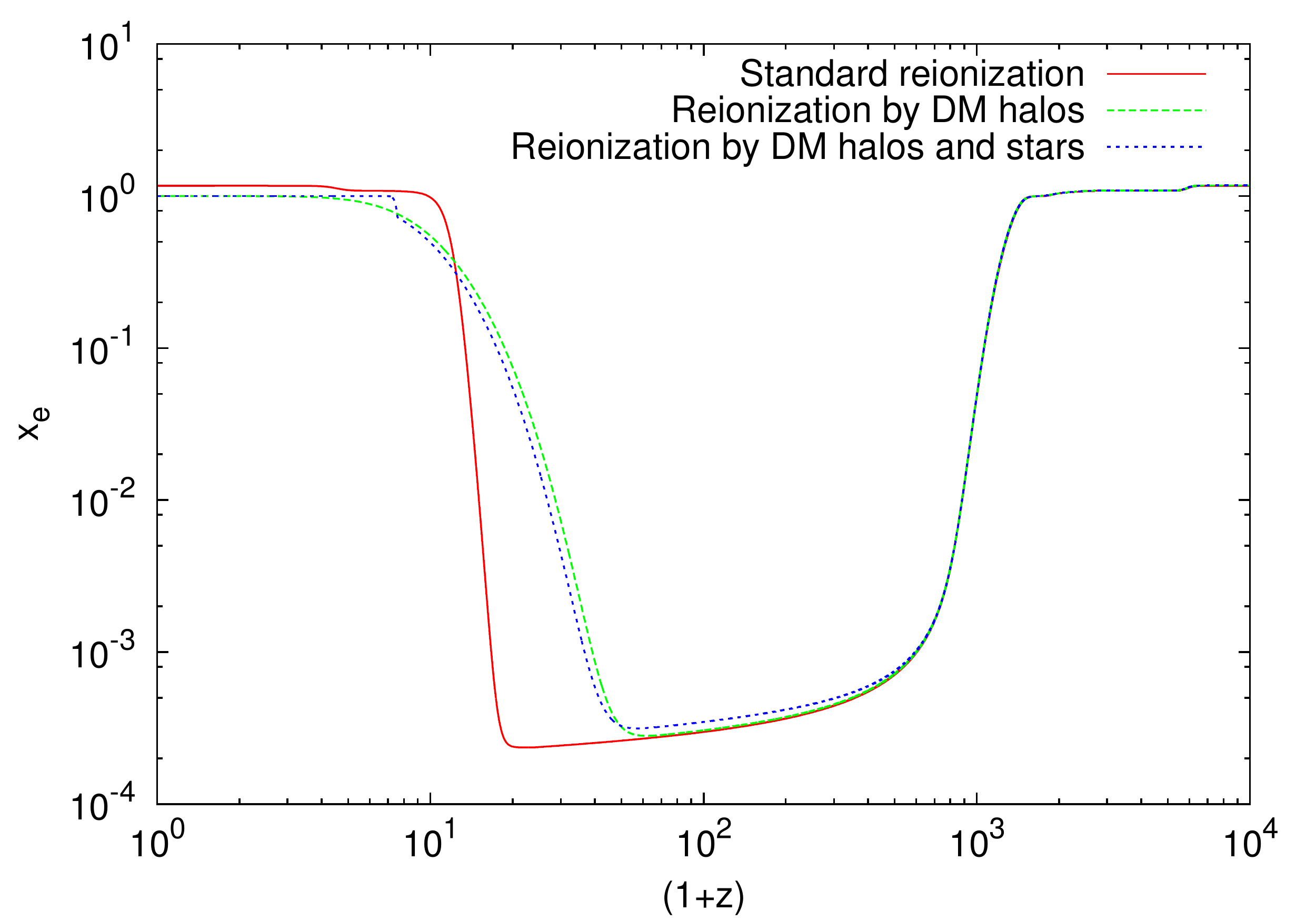}
\caption{Free electron fraction for the three best-fitting models assuming reionization from stars (with the usual single step parameterization), from DM annihilation in  halos, or from both with an additional Gunn-Peterson prior. All curves were obtained using {\sc hyrec} in mode {\tt RECFAST}.
\label{halos_mixed_xe}}}

\FIGURE{
\includegraphics[scale=0.30]{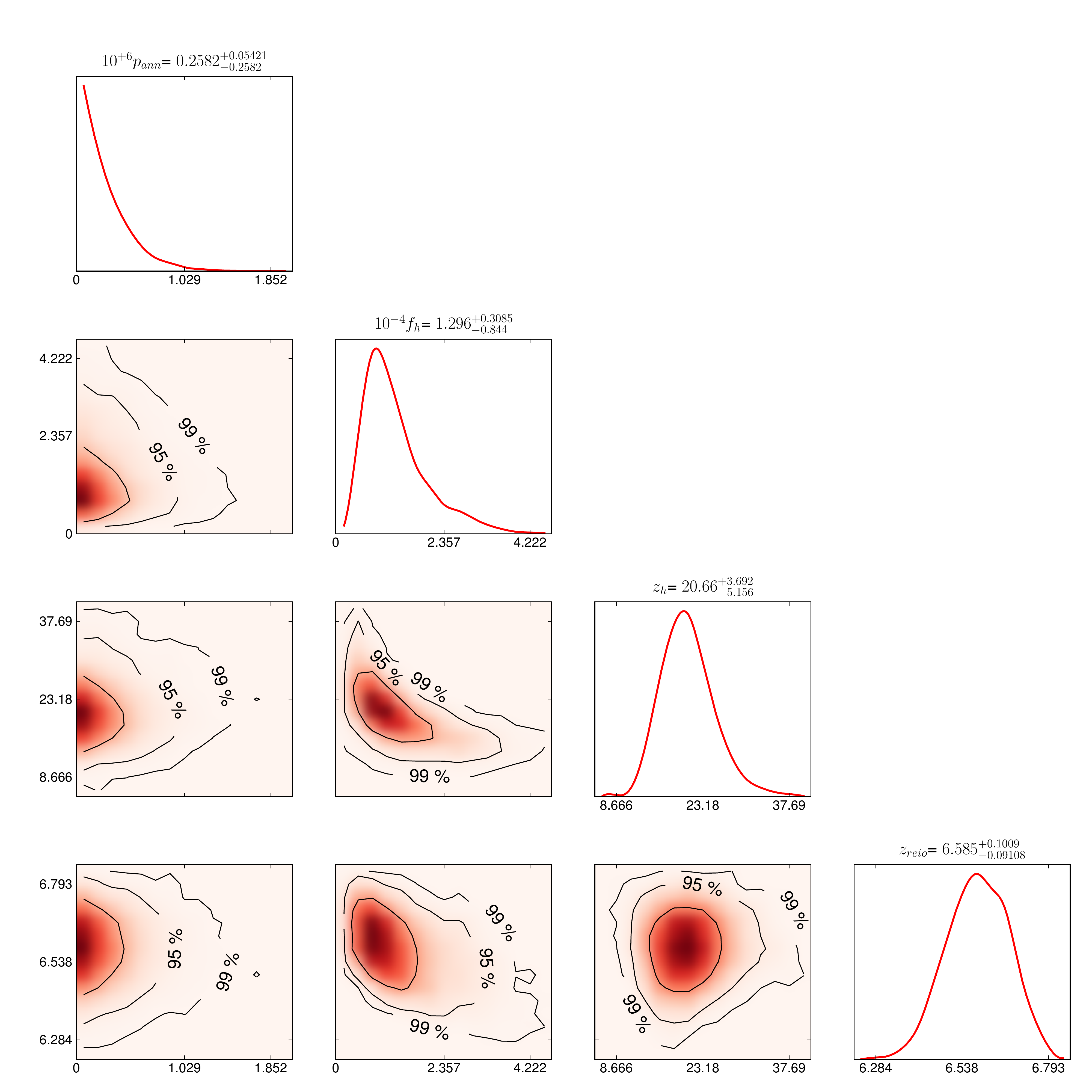}
\caption{Assuming a $\Lambda$CDM model with free effective neutrino number, dark matter annihilation (background and halos) and single-step reionization from stars, marginalized probability distribution of the annihilation parameters $p_{\rm ann}$, $f_{\rm h}$, $z_{\rm h}$ and $z_{\rm reio}$ given WMAP7 and SPT data and an additional Gunn-Peterson prior.
\label{mixed_monte}}}

Our results are summarized in the last column of Table~\ref{table1} and in the triangle plot of figure~\ref{mixed_monte}. With respect to the previous model of section~\ref{reio_halo}, we have one more parameter $z_{\rm reio}$, that is very constrained by the Gunn-Peterson prior: it can only fluctuate in the range 
$6.4<z_{\rm reio}<6.8$ (95\% C.L.). The posterior probability of $f_{\rm h}$ and $z_{\rm h}$ are shifted to slightly smaller values, since DM annihilation in halos is only expected to contribute to a fraction of the optical depth. For all other parameters, the results are essentially identical to those of the previous case. The minimum effective chi square is also unchanged. The discussion of section~\ref{reio_halo} concerning $f_{\rm h}$ and $z_{\rm h}$ still applies: $z_{\rm h}$ is fixed to a range that coincides with expectations from structure formation, and values of  $f_{\rm NFW}(c_{\rm h})$ in the range from $10^3$ to $10^5$ could be accommodated provided that $p_{\rm ann}$ and $f_{\rm NFW}(c_{\rm h})$ fulfill the relation (\ref{rel_pann_f}) with $f_{\rm h}$ in the range $2100<f_{\rm h}<28600$ (95\% C.L.).

In conclusion of this subsection, we see that this mixed model for reionization is interesting: DM annihilation in halos could explain the value of the optical depth probed by CMB data, while reionization from star formation at $z\simeq 6.5$ would complete the reionization process and explain Gunn-Peterson observations.

\subsection{Including an upper bound on the IGM temperature\label{tigm}}

The best-fitting models of sections~\ref{reio_halo} and \ref{gunn_peterson} have a halo parameter $f_{\rm h}$ of the order of $10^4\,{\rm m}^3/{\rm s}/{\rm kg}$, leading to a matter temperature of the order of $10^5$K at low redshift (see figure~\ref{halos_xe}). This estimate of the average matter temperature in the universe should be taken with a grain of salt, since we did not account for inhomogeneities in the matter distribution, nor for the thermodynamical evolution of the inter-galactic medium (IGM) during structure and star formation.

However, Lyman-$\alpha$ observations suggest that the IGM temperature is of the order of a few times $10^4$K in the redshift range $2 \leq z \leq 4.5$. Reference \cite{Cirelli2009} pointed out that these measurements should provide at least an upper bound on the average temperature enhancement due to DM annihilation. 

In other words, the results of the previous two subsections are compatible with Lyman-$\alpha$ observations only if we are modelling the temperature evolution incorrectly. The ansatz that a fraction $(1+2x_e)/3$ of the energy injected into the gas by DM annihilation goes into heating might be incorrect at low redshift; or the IGM temperature growth might be limited by some temperature regulation mechanisms not described by our simplistic set of equations (such as, for instance, line cooling or Bremsstrahlung effects). If instead our temperature evolution law is realistic, then DM annihilation cannot explain the reionization of the universe alone, and cannot even contribute sufficiently to reionization at $z\sim6$ in order to explain Gunn-Peterson bounds with a mixed reionization model, based on annihilation plus a single-step star formation process.

\FIGURE{
\includegraphics[scale=0.30]{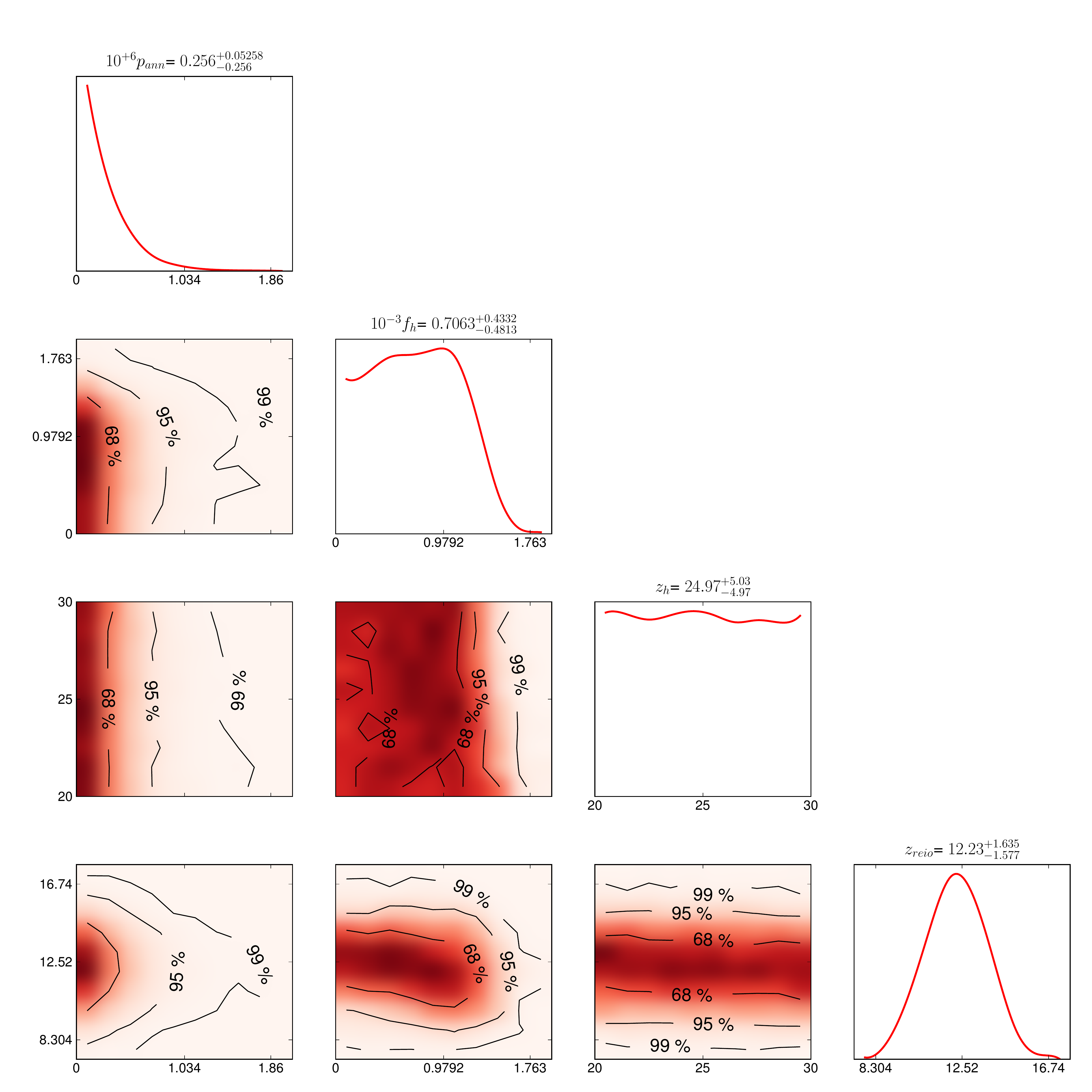}
\caption{Assuming a $\Lambda$CDM model with free affective neutrino number, dark matter annihilation (background and halos) and single-step reionization from stars, marginalized probability distribution of the annihilation parameters $p_{\rm ann}$, $f_{\rm h}$, $z_{\rm h}$ and $z_{\rm reio}$ given WMAP7 and SPT data, a prior $20 \leq z_{\rm h}\leq 30$ and an upper bound on the IGM temperature.
\label{halo_reio_tigm}}}

 It is still interesting to perform a parameter extraction with a prior on $T_{\rm M}$, while assuming a mixed reionization from halos and stars, in order to check whether IGM temperature estimates provide a stronger bound on $p_{\rm ann}$ than the CMB alone.
In this section, we will repeat our analysis with a conservative upper bound on the IGM temperature at low redshift, inspired from figure 6 (left) in \cite{Schaye:1999vr}:
\begin{align}
T_{\rm M}(z=2)  \leq 3.2\times10^{4} {\rm K}.
\end{align}
We implemented this constraint in the form of a top-hat prior in {\sc montepython}. We checked in presence of such a prior, the model of section \ref{reio_halo} in which reionization is caused entirely by DM annihilation requires unrealistically high values of $z_h\sim 100$, totally incompatible with structure formation models. 

Like in the previous subsection, we implemented star reionization into {\sc class} ``by hand'': below some arbitrary redshift $z_{\rm reio}$, $x_e(z)$ is cut and matched continuously to a half-hyperbolic tangent centered on $z_{\rm reio}$, reaching an asymptotic value of one for $z \rightarrow 0$. We kept the transition width parameter at its default value: $\delta z = 1.5$. We also imposed a top-hat prior $20\leq z_{\rm h} \leq 30$ in order to ensure that halos form at a realistic redshift, compatible with simulations of structure formation.

Our results are summarized in the last column of Table~\ref{table1} and in the triangle plot of figure~\ref{halo_reio_tigm}. The upper bound on $f_{\rm h}$ is reduced by one order of magnitude due to the IGM temperature constrain: the 95\%CL upper limit on $f_{\rm h}$ decreases from $f_{\rm h}^{\rm max}=25600\,{\rm m}^3/{\rm s}/{\rm kg}$ to $1400\,{\rm m}^3/{\rm s}/{\rm kg}$. This leads to a stronger bound on $p_{\rm ann}$:
\begin{equation}
p_{\rm ann}(z=600) < 0.05 \times 10^{-6} \left[ \frac{p_{\rm ann}(600)}{0.5 p_{\rm ann}(0)}\frac{61^3}{(1+z_{\rm F})^3} \frac{10^4}{f_{\rm NFW}(c_{\rm h})}\right] {\rm m}^3/{\rm s}/{\rm kg}~~(95\% \textrm{C.L.})
\label{tigm_constraint}
\end{equation}
Assuming that the factor between brackets is equal to one, this bound is almost twenty times stronger than the one inferred from annihilation in the smooth component only.

The heating effect of DM annihilation may also enhance the kinetic Sunyaev-Zel'dovich effect \cite{Sunyaev1972}, leave a signature in secondary CMB anisotropies, and provide a further test of this model \cite{Natarajan2012}. We do not study this aspect in our work.

\section{Conclusion and Outlook}

We studied different possible contributions of annihilating Dark Matter to the thermal history of the universe. We
confirmed previous results that the annihilation of the background DM distribution has non-trivial effects on the CMB, leading to the constraint
\begin{equation}
p_{\rm ann}(z \sim 600)< 0.91 \times 10^{-6}{\rm m}^3/{\rm s}/{\rm kg} \qquad {\rm (WMAP7+SPT, ~95\% C.L)},\label{bound_back}
\end{equation}
with a negligible impact of the variations of $p_{\rm ann}(z)$ in the range $100<z<2500$ suggested by a realistic study of DM annihilation channels.

We also showed that DM annihilation in halos could explain entirely the reionization of the universe from the point of view of CMB observations. In addition, if the constraints $x_{\rm HI}(6)\geq10^{-3}$, $x_{\rm HI}(5.5)\leq 10^{-4}$ inferred from the Gunn-Peterson effect hold, and if we assume that reionization from stars takes place abruptly in one step, then a mixed model with DM annihilation in halos and star formation at $z \simeq 6.5$ could explain simultaneously CMB observations and the above bounds. However, these models tend to reheat the IGM well above the typical temperatures indicated by Lyman-$\alpha$ observations, unless our modeling of the matter temperature evolution at low redshift is incorrect.

Our most important conclusion is that constraints on DM annihilation in
halos tend to be stronger than those from the smooth background distribution of
DM, especially if we include a realistic upper bound on the matter temperature at low redhsift. Assuming Press-Schechter theory and NFW profiles, we see from eq.~(\ref{tigm_constraint}) that for $f_{\rm NFW}(c_{\rm h})=10^3$, $z_{\rm F}=60$ and $[p_{\rm ann}(600)/p_{\rm ann}(0)]=5$, the constraint coming from halos and from the smooth background are comparable. If in reality halos are more concentrated than in this simple model, then constraints on $p_{\rm ann}$ from annihilation in halos superseed those from annihilation in the background.
We summarize our constraints on $p_{\rm ann}$ and their implications for the DM mass and cross-section in figure~\ref{fig_sigma}. A WIMP with standard thermal cross-section $\langle \sigma v \rangle \simeq 3 \times 10^{-26}\,$cm$^3/$s is constrained by annihilations in the smooth background to have a mass larger than $18 \times f(z=600)$~GeV$/c^2$ (95\% C. L.). According to our simple model for DM annihilation in halos, the bound increases to about $100[\frac{f_{\rm NFW}}{10^4}]f(0)$~GeV$/c^2$ in order to avoid reionizing the universe too early (or even $1700[\frac{f_{\rm NFW}}{10^4}]f(0)$~GeV$/c^2$ when including IGM temperature bounds).
\\
\FIGURE{
\includegraphics[width=7cm,angle=-90]{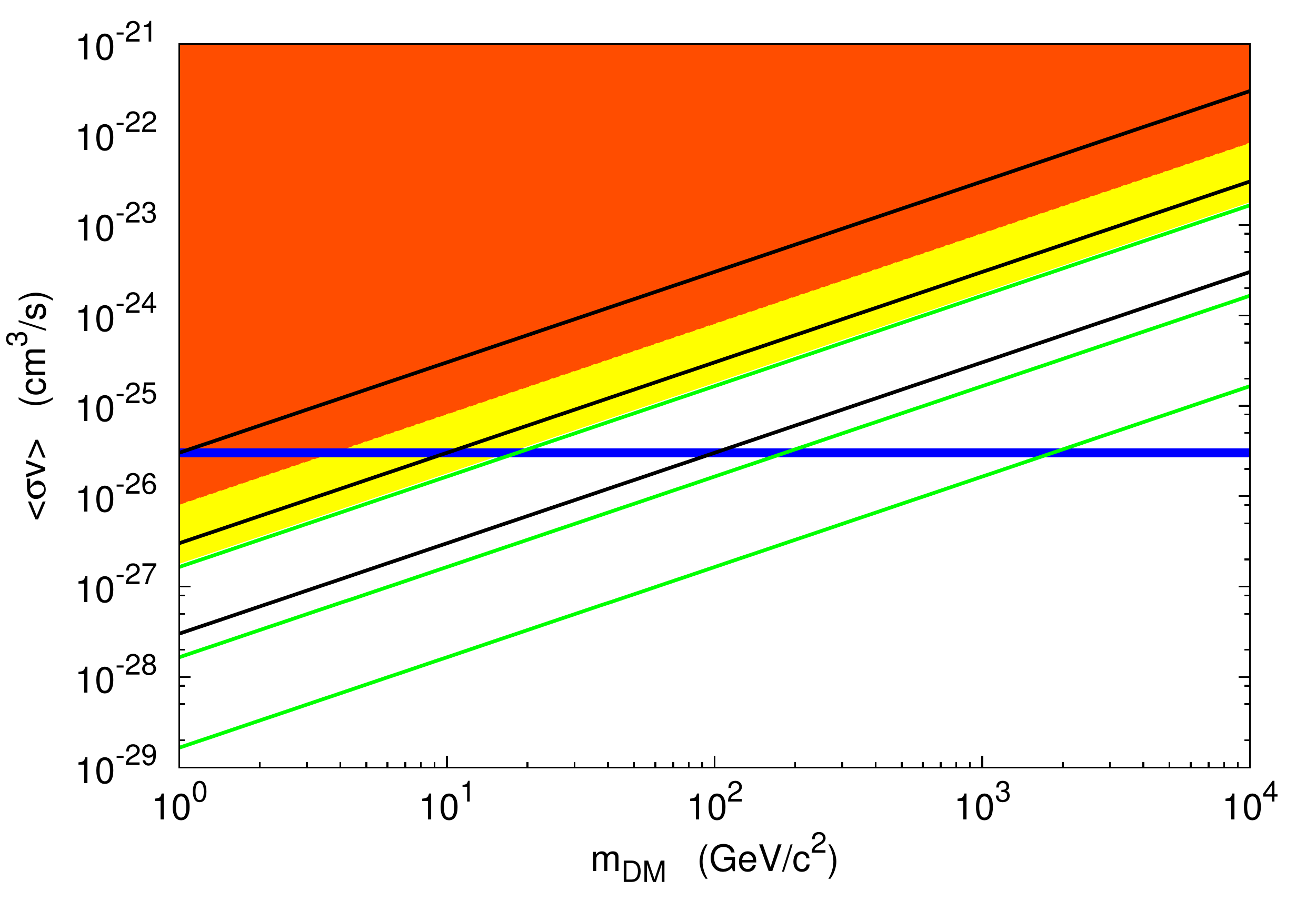}
\caption{\label{fig_sigma} Summary of our bounds on $p_{\rm ann}$ translated into constraints for the DM annihilation thermal cross-section $\langle \sigma v \rangle$ and mass $m_{\rm DM}$. Points in the shaded regions are above the 95\% preferred region for $p_{\rm ann}(z=600)$, considering only annihilation in the smooth DM background, and assuming either $f(z=600)=0.2$ (dark shade) or $f(z=600)=0.9$ (light shade): these two extreme assumptions cover the plausible range for $f(z=600)$ in the case of WIMP annihilation, see~\cite{Slatyer2009}. The three black lines correspond to the CMB bounds inferred from DM annihilation in halos, assuming $f_{\rm NFW}(c_{\rm h})=10^3$ (top), $10^4$ (middle) or $10^5$ (bottom), and taking in all three cases $z_{\rm F}=60$ and $f(z \simeq 0)=0.1$ (or in other words, $f(z=600)=0.5$ and $[p_{\rm ann}(600)/p_{\rm ann}(0)]=5$). When a realistic upper bound on the matter temperature at low redshift is taken into account, the bounds move to the green lines. The horizontal lines shows the standard WIMP thermal cross-section.}}

This work contains a systematic analysis of DM annihilation in halos, where values of unknown parameters (including those describing structure formation) are freely varied and fitted to the data. Several authors have previously investigated the effect of DM annihilation in halos, for particular models with fixed parameter values. The authors of \cite{Belikov2009,Cirelli2009,Hutsi:2011vx} reached the conclusion that annihilation in halos is usually inefficient. Indeed, they choose some DM density profiles corresponding roughly to $f_{\rm NFW}=100$ or 400. We found that halo bounds become stronger than smooth background bounds only for $f_{\rm NFW}>4200$. Hence our results are not contradicting these previous works. 

Many aspects of this analysis could be improved. For example, we
neglected Helium throughout the whole discussion. The effect of Helium has been
studied by Galli et al. \cite{Galli2011} (neglecting halo effects) and does not change the result
significantly. 
The energy fraction going into ionization $\chi_i$ and Lyman-$\alpha$ excitation $\chi_\alpha$ was also approximated, motivated by a common sense argument by Chen and Kamionkowski \cite{Chen:2003gz}. However, the exact behavior of these quantities have a negligible effect on the CMB.
The average DM density during non-linear structure formation has been approximated with a basic Press-Schechter model and NFW profiles. We could have imposed priors on the parameters of this model inferred from N-body simulations, or tried different profiles (Einasto profile, etc.), or a more realistic differential mass function \cite{Sheth1999}. Instead of the  Press-Schechter model, we could have accounted for halo formation using the excursion set formalism~\cite{Maggiore:2009rv}.  One could try to model the matter temperature evolution at low redshift more accurately, taking into account matter inhomogeneities and the complicated thermodynamical evolution of the IGM (including, for instance,  line cooling or Bremsstrahlung effects). However, all these refinements are probably unnecessary at the moment, given the large error bars on the optical depth inferred from CMB observations.

Throughout this work, we assumed that DM annihilates. A similar study can be performed in the case of decaying dark matter \cite{Chen:2003gz,Zhang:2007zzh,Kasuya:2006fq,Yeung:2012ya}. In that case, the energy injection rate varies  like $\bar{\rho}_{\rm DM}$ (instead of $\bar{\rho}_{\rm DM}^2$), i.e. like $(1+z)^3$. Hence, the effect of DM decay in the smooth DM background is not very different from the effect of DM annihilation in halos, studied in section~\ref{sec:halos}. Note however that for a wide range of masses, constraints on the DM lifetime inferred from current CMB observations are not as strong as those inferred from cosmic rays \cite{Bell:2010fk,Cirelli:2012ut}.

In a few months from now, results from the Planck satellite data may lead to a significant improvement of these bounds, and bring complementary information on the DM mass and cross-section (or lifetime) with respect to direct and other indirect DM search.

\section*{Acknowledgements}

We would like to thank Alexey Boyarsky, Marco Cirelli,  Silvia Galli, Oleg Ruchayskiy and Pasquale Serpico for enlightening discussions and detailed comments on this manuscript.
This project is supported by a research grant from the Swiss National Science
Foundation. Y.~A.-H. is supported by the National Science Foundation
grant number AST-080744 and acknowledges support from the Max Planck
Institute for Astrophysics during the month of July 2012.

\section*{Appendix A: modifications in {\sc hyrec}}
\label{appendix_hyrec}

In {\sc hyrec}, the evolution equation are written in function of time for $x_e$ and of $\ln a$ for $T_{\rm M}$. In addition, the units are CGS+eV for temperatures, except in the two functions describing the temperature evolution, where we have Kelvin. So, the equations are
\begin{align}
H\frac{d x_e}{d \ln a}=&\frac{1}{H}\left[C\left(-n_H x_e^2 \alpha_B+\beta_B (1-x_e) e^{-\frac{E_{21}}{T_r}}\right)\right. \label{hyrec_xe}\nonumber
\\ &\left.+\rho_{\rm c}^2c^2\Omega_{\rm DM}^2\frac{1-x_e}{3n_H}(1+z)^6p_{\rm ann}(z)\left(\frac{1}{E_i}+\frac{1-C}{E_\alpha}\right)\right],
\\ \frac{d T_{\rm M}}{d \ln a}=&-2T_{\rm M} +\frac{1}{H}\left[\frac{8\sigma_T a_r T_r^4}{3 m_e c}\frac{x_e}{1+x_e+f_{He}} (T_r-T_{\rm M})\right. \nonumber
\\& +\left.\frac{2}{3 k_b}\frac{1+2x_e}{3n_H}\frac{1}{1+x_e+f_{He}} \rho_{\rm c}^2c^2\Omega_{\rm DM}^2(1+z)^6p_{\rm ann}(z)\right].
\end{align}
These equations are equivalent with the ones in {\sc recfast}, but the $C$-factor is now defined as:
\begin{equation}
C=\frac{\frac{3}{4}R_{Ly\alpha}+\frac{1}{4}\Lambda_{2s,1s}}{\beta_B+\frac{3}{4}R_{Ly\alpha}+\frac{1}{4}\Lambda_{1s,2s}},
\end{equation}
with the two-photon rate $\Lambda_{2s,1s}=8.22458$ s$^{-1}$ and the escape rate of Lyman-$\alpha$ photons $R_{Ly\alpha}=\frac{8\pi H}{3 n_H (1-x_e) \lambda_{Ly\alpha}^3}$. 

\subsubsection*{Quasi steady-state equation}
In {\sc hyrec}, we can find a function describing the temperature evolution in the quasi steady-state approximation. In general, as seen above, the equation for the temperature in the presence of annihilating Dark Matter is
\begin{equation}
\frac{d T_{\rm M}}{d \ln a}=-2 T_{\rm M} +\gamma(T_r-T_{\rm M})+\left. \frac{d T_{\rm M}}{d \ln a}\right|_{\rm DM}.
\label{temp}
\end{equation}
The quasi steady-state approximation consists of considering the case when the second term in equation (\ref{temp}) is bigger than the other two, i.e. when $\gamma \gg 1$. In this situation, $T_{\rm M}\approx T_r$ and $\frac{d T_{\rm M}}{d \ln a} \approx -T_{\rm M}$,  thus
\begin{equation}
T_{\rm M}\approx \frac{T_r}{1+\gamma^{-1}}+\gamma^{-1}\left. \frac{d T_{\rm M}}{d \ln a}\right|_{\rm DM}.
\end{equation}

\subsubsection*{The different modes of {\sc hyrec}}

The above evolution equation for $x_e$ is used when {\sc hyrec} runs in the modes {\tt peebles} or {\tt recfast}. For the {\tt EMLA} mode, there exist two different $C$-factors, namely $C_{2s}$ and $C_{2p}$
\begin{align}
C_{2s}&=\frac{\Lambda_{2s1s}+R_{2s\rightarrow 2p}\frac{R_{Ly\alpha}}{\Gamma_{2p}}}{\Gamma_{2s}-R_{2s\rightarrow2p} \frac{R_{2p\rightarrow 2s}}{\Gamma_{2p}}},
\\ C_{2p}&=\frac{R_{Ly\alpha}+R_{2p\rightarrow 2s}\frac{\Lambda_{2s1s}}{\Gamma_{2s}}}{\Gamma_{2p}-R_{2p\rightarrow 2s} \frac{R_{2s\rightarrow 2p}}{\Gamma_{2s}}}.
\end{align}
The inverse life times are defined by:
\begin{align}
\Gamma_{2s}&=B_{2s}+R_{2s \rightarrow 2p}+\Lambda_{2s,1s},
\\ \Gamma_{2p}&=B_{2p}+R_{2p \rightarrow 2s}+R_{Ly\alpha},
\end{align}
where $B_i$ are the ionization coefficient and $R_{i\rightarrow j}$ the transition coefficients. We can take $ R_{2s \rightarrow 2p}=3 R_{2p \rightarrow 2s}$ since there are 3 times more states in 2p than in 2s.
\\$C_{2s}$ ($C_{2p}$) represents the probability that a hydrogen atom
initially in the 2s (2p) state reaches the ground sate before being
ionized.  The Lyman-$\alpha$ line is the excitation from 1s to 2p. So
the new factor in equation (\ref{hyrec_xe}) should be\footnote{We used
$C_{2p}$ assuming that excitations were mostly $1s\rightarrow 2p$, as
would be the case if DM annihilations lead to additional Ly-$\alpha$
photons. If excitations are instead collisional, this treatment is not
formally valid; however, this would represent a correction to a
process that is already subdominant and we need not worry about such
subtleties here.} $C=C_{2p}$. Exactly the same approach is used for the {\tt full} mode of {\sc hyrec}.

\bibliographystyle{utcaps}
%\bibliography{annihil.bbl}

\end{document}